\documentclass[11pt]{article}

\usepackage{times}
\usepackage{amsmath}
\usepackage{amsthm}
\usepackage{graphicx,epsfig}

\newcommand {\E}{\mathbf E}

\newtheorem{theorem}{Theorem}[section]

\newtheorem{lemma}[theorem]{Lemma}
\newtheorem{claim}[theorem]{Claim}
\newtheorem{corollary}[theorem]{Corollary}

\newtheorem{define}{Definition}

\newtheorem{ex}{Example}[section]

\newcommand{\eat}[1]{}
\renewcommand{\paragraph}[1]{\medskip \noindent {\bf{#1}}}

\newtheorem{assumption}{Assumption}

\advance \topmargin by -\headheight
\advance \topmargin by -\headsep
\evensidemargin \oddsidemargin
\begin{document}
\title{Incentive Compatible Budget Elicitation in Multi-unit Auctions}
\author{Sayan Bhattacharya\thanks{Corresponding author.}  \qquad Vincent
  Conitzer\thanks{Supported by an Alfred P. Sloan Research
Fellowship and by  NSF grant IIS-0812113.} \qquad Kamesh Munagala\thanks{Supported by an Alfred P. Sloan
    Research Fellowship, and by NSF via a CAREER award and grant
    CNS-0540347.} \qquad Lirong Xia\thanks{Supported by a James B. Duke Fellowship and NSF grant IIS-0812113.}  \\ \\ 
 Department of Computer Science, Duke University\\ 
 {\small \tt \{bsayan,conitzer,kamesh,lxia\}@cs.duke.edu} }
 \date{}
 \maketitle

 \begin{abstract}
   In this paper, we consider the problem of designing incentive compatible
   auctions for multiple (homogeneous) units of a good, when bidders have  private
   valuations and private budget constraints.    When only the valuations are private
   and the budgets are public, Dobzinski {\em et al}~\cite{nisan} show that
   the {\em adaptive clinching} auction is the unique incentive-compatible
   auction achieving Pareto-optimality. They further show that this auction is not truthful with private budgets, so that there is no    deterministic  Pareto-optimal auction with private budgets. Our
   main contribution is to show the following Budget Monotonicity property
   of this auction: When there is only one infinitely divisible good, a
   bidder cannot improve her utility by reporting a budget smaller than the
   truth.  This implies that the adaptive clinching auction is incentive compatible when over-reporting the budget is not possible (for instance, when funds must be shown upfront). We can also make reporting larger budgets suboptimal with a small randomized modification to the auction. In either case, this makes the modified auction Pareto-optimal with private budgets. We also show that
   the Budget Monotonicity property does {\em not} hold for auctioning indivisible units of the good,
   showing a sharp contrast between the divisible and indivisible cases.
 
   The Budget Monotonicity property also implies other improved  results in this context. For revenue maximization, the same auction improves the
   best-known competitive ratio due to Abrams~\cite{abrams} by a factor of
   $4$, and asymptotically approaches the performance of the optimal
   single-price auction.

   Finally, we consider the problem of revenue maximization (or social
   welfare) in a Bayesian setting. We allow the bidders have public size constraints (on the amount of
   good they are willing to buy) in addition to private budget
   constraints. We show a simple poly-time
   computable $5.83$-approximation to the optimal Bayesian incentive
   compatible mechanism, that is implementable in dominant strategies. Our technique again crucially needs the ability to prevent bidders from over-reporting budgets via randomization.  We
   show the approximation result via designing a rounding scheme for an LP relaxation of the problem (related to Myerson's LP), which may be
   of independent interest.
\end{abstract}

\thispagestyle{empty}
\newpage
\setcounter{page}{1}

\section{Introduction}
In this paper, we consider the problem of designing incentive compatible
auctions for multiple homogeneous units of a good. This problem has received
significant attention in the non-Bayesian setting starting with the work of
Goldberg {\em et al}~\cite{Goldberg04}. We focus on the scenario where
bidders not only have a private valuation per unit of the good, but also a
private budget, that is the total amount of money they are able to pay. The budget constraint is {\em hard}; a bidder gets a utility of negative infinity if she has to pay a total price larger than her budget. In
this model, the natural problems to consider are maximizing social
welfare and the auctioneer's revenue. Both these aspects have been
considered in previous work~\cite{abrams,borgs,nisan} in an adversarial
setting.

The key difficulty with budget constraints is that the utilities are no longer quasi-linear. This makes
mechanisms such as VCG no longer applicable.  Based on the random
partitioning framework of Goldberg {\em et al}~\cite{Goldberg04}, Borgs
{\em et al}~\cite{borgs} present a truthful auction whose revenue is
asymptotically optimal compared to that of the optimal single-price
mechanism. Using the same framework, Abrams~\cite{abrams} gives a different auction that improves  this result for a
 range of parameters (but is not asymptotically optimal). 

More recently, Dobzinski {\em et al}~\cite{nisan} presented the {\em adaptive clinching auction} based on the clinching auction of Ausubel~\cite{ausubel}. This is an ascending price auction where each bidder maintains a demand, which is the amount of item she is willing to buy given the current price and her residual budget. Initially, the demand is larger than supply. If the total demand of the remaining bidders is less than the supply of items, the bidder {\em clinches} the difference at the current price. The bidder drops out of the auction if the price exceeds her valuation, and the auction stops when the total demand falls below total supply. Though the auction rules seem simple, it still defines a differential process for auctioning an infinitely divisible good with no closed form solution, except in special cases.

It is not difficult to show that this auction is incentive compatible when the budget constraints are public knowledge.  Dobzinski {\em et al}~\cite{nisan}  show that in the public budget setting, it is the {\em only}  such auction that is Pareto-optimal (PO), meaning that no pair of agents (including the auctioneer) can simultaneously improve their utilities by trading with each other\footnote{In this setting, no truthful auction can maximize social welfare~\cite{borgs}, hence the focus on Pareto-optimality.}. They further show that with public budgets, this auction has better revenue properties  than the auctions in~\cite{abrams} and~\cite{borgs}: It improves the former by a factor of $4$, and like the latter, is asymptotically optimal.  However, their main result is negative: This auction is not truthful when the budgets are private knowledge, so that there is no Pareto-optimal truthful auction in this case.

\medskip
\noindent{\bf Our Results.}  The negative result in~\cite{nisan} holds for any auction with private budgets that needs to satisfy three properties
ex-post\footnote{In this paper, {\em ex-post} will mean the property holds for randomized mechanisms regardless of the outcome of randomization.}: Voluntary participation (VP), Incentive compatibility (IC), and No
positive transfers (NPT). These properties are standard, and defined in Section~\ref{sec_intro}. The main result in this paper
is to show that there is indeed a Pareto-optimal {\em randomized} mechanism
with private budgets. Here, the prices and quantities are random variables; the (IC) and (VP) properties are satisfied
in {\em expectation} over these random variables; and (NPT) and (PO) are satisfied ex-post\footnote{In the case of one infinitely divisible good, the auction also needs to be {\em anonymous} for the negative result in~\cite{nisan} to hold, meaning that the auction is symmetric for bidders with identical types; our 
randomization can also easily be made anonymous.}.   The key to showing this 
result is to develop a novel structural characterization of the adaptive
clinching auction in the case of one infinitely divisible good.

\medskip\noindent{\bf Budget Monotonicity and Randomization.} We show an
intuitive property of the adaptive clinching auction in the case of divisible goods: A bidder cannot gain utility by
reporting budget lower than the truth. We term this property 
  Budget Monotonicity. Though this property seems simple, there is no
reason to assume it holds: {\em In fact, this property is  false for the
adaptive clinching auction in the case of indivisible units} (Theorem~\ref{thm:eg}).The major difficulty in the proof is that the adaptive clinching auction continuously makes allocations at different prices, so that the utility is a complicated function of all the budgets and valuations. In fact, an analysis of this auction is left as
an open question\footnote{Since the focus of~\cite{nisan} is to prove uniqueness and impossibility, they mainly analyze the auction for two bidders with carefully chosen valuations and budgets. In contrast, we need to develop characterizations for the general case.} in~\cite{nisan}.  We show this result
by carefully coupling the behavior of two auctions that differ only in the
reported budget of one bidder. The proof also establishes several structural results
about this auction that are of independent interest.

\medskip
Budget Monotonicity for an infinitely divisible good implies Pareto-optimality fairly directly, since all we need to prevent is a bidder over-reporting her budget. We can do this in several ways, the simplest being randomization. In the  Randomized Extraction Scheme (see Section~\ref{sec_intro}), the mechanism simply extracts the whole budget or zero price so that the expected price extracted is equal to the price charged by the deterministic auction. Therefore, if a bidder gets nonzero allocation by over-reporting her budget, then with non-zero probability, she pays her reported budget and her expected utility is $-\infty$. This scheme can be applied to any deterministic auction to prevent reporting larger budgets than the truth. We show in Section~\ref{sec_intro} that the randomized version preserves Pareto-optimality ex-post, and maintains (IC) and (VP) in expectation if the deterministic auction was monotone (and charges non-zero price for non-zero allocations).

 The monotonicity result for the adaptive clinching auction holds only for
one infinitely divisible good. In the case of finitely many indivisible
units, we simply run the adaptive clinching auction assuming one infinitely
divisible good, and perform a randomized allocation in the end (Corollary~\ref{cor:manygoods}; also see~\cite{abrams,borgs}). The resulting auction is (VP) and (IC) in expectation,  and is also
Pareto-optimal. 

\medskip
Though our mechanism is randomized, the randomness introduced in the price is quite small: {\em It affects the price charged to only one bidder} (Lemma~\ref{lem:onlyone}). However, for the randomization to be (IC), we crucially need the assumption that the utility of a bidder for paying more than her true budget is $-\infty$. For smoother utility functions, the   Budget Monotonicity property can be used in other ways to make the deterministic auction itself truthful and Pareto-optimal: For instance, a standard assumption in spectrum auctions~\cite{pino} is that the bidder can be forced to show ``proof of funds" for her reported budget (for instance, a bank statement), and this prevents her from over-reporting  her budget regardless of her utility function. 

\medskip
\noindent{\bf Revenue Properties.} As another consequence of Budget Monotonicity, the improved  revenue properties of the adaptive clinching auction over the auctions in~\cite{abrams,borgs} in the case of public budgets (shown in~\cite{nisan}) carry over to the randomized version of the auction even with private budgets, and hence, this auction improves the competitive ratio in~\cite{abrams} by a factor of $4$, and like the auction in~\cite{borgs}, is asymptotically optimal.

\medskip\noindent{\bf  Bayesian Setting.}  In this setting, the auctioneer maintains independent discrete distributions on the possible valuations and budgets for each bidder, and is interested in designing a poly-time computable mechanism for optimizing 
expected revenue (resp. social welfare).  In this setting, we allow the  bidders to have a public size constraint on the amount of item they can buy in addition to a private valuation and private budget. 

Variants of this model have been considered before~\cite{myerson,vohra,laffont,maskin,shuchi,timr}, and the optimal solution can indeed be encoded as an (exponential size) linear program. The key challenge now becomes designing  polynomial time computable mechanisms. It is well-known~\cite{myerson,vohra} that the optimal mechanism has a simple structure related to the VCG mechanism in the case of {\em  i.i.d.} distributions and no size constraints. It is unlikely that such a structure holds in the general setting, and we instead consider designing approximately optimal mechanisms. The budget constraints however make a poly-time relaxation of the problem non-linear. However, if we only encode that utility decreases for under-reporting budgets, the program becomes linear;  we again use randomization to prevent over-reporting budgets.  Using this, we show a poly-size linear program relaxation with a rounding scheme that yields a $5.83$ approximation to the optimal Bayesian IC mechanism  when the type space is discrete; this mechanism is  implementable in dominant strategies. This rounding technique may be applicable in other related scenarios.

\medskip\noindent{\bf Organization of the Paper.} In Section~\ref{sec_intro}, we
define the notions of truthfulness and Budget Monotonicity. In
Section~\ref{sec:clinch}, we describe the adaptive clinching auction and
show some basic properties in the infinitely divisible good case. In
Section~\ref{sec_proof}, we sketch the proof of Budget Monotonicity of this auction,
which implies that the randomized version satisfies Pareto-optimality with
private budgets. The proof of this claim very technical, and is hence presented in its entirety in Appendix~\ref{app:monotone}. In Section~\ref{app:size}, we consider the Bayesian setting and show an LP rounding scheme that achieves a $5.83$ approximation to the expected revenue (resp.
social welfare) even when bidders have public size constraints.

\section{Preliminaries}
\label{sec_intro}
We will mainly consider the case when there is one unit of infinitely
divisible good and $n$ bidders. Bidder $i$ has a private valuation $\eta_i$
per unit quantity, and private budget $\beta_i$. Suppose bidder $i$ reports
valuation and budget $(v_i, B_i)$. The auction is a (randomized) mechanism that (probabilistically) maps the $(\vec{v}, \vec{B})$ into a quantity $X_i$ the bidder obtains and a total
price $P_i$ the bidder pays; note that these quantities are allowed to be random variables in this paper. Since there is one unit of the good, we have $\sum_i X_i \le 1$.

The only difference in the case of auctioning $m$ indivisible
copies of the good is that in this case, $X_i \in \{0,1,2,\ldots,m\}$ and
$\sum_i X_i = m$.  (The infinitely divisible good case is the limit when $m \rightarrow \infty$.)  In the
subsequent discussion we will assume one infinitely divisible good unless
otherwise stated.

Let $v_{-i}, B_{-i}$ denote the reported valuations and budgets of bidders
other than $i$.

Bidder $i$ has the following utility function: If $P_i > \beta_i$, then
his utility is $-\infty$: this corresponds to the total price exceeding his
budget. If $P_i \le \beta_i$, then his utility is $u_i = \eta_i X_i - P_i$.

The goal is to design a randomized auction that
satisfies the following four properties. Note that in the (VP) and (IC) conditions, the
expectation is over the randomness introduced by the mechanism. 

\begin{description}
\item[Voluntary Participation (VP):] If $v_i = \eta_i$ and $B_i = \beta_i$,
  then regardless of $v_{-i}, B_{-i}$, we have $\E[u_i] \ge 0$.
\item[Incentive Compatibility (IC):] Regardless of $v_{-i}, B_{-i}$,
  $\E[u_i]$ is maximized when $v_i = \eta_i$ and $B_i = \beta_i$.
  \item[No Positive Transfers (NPT):] Regardless of $\vec{v}, \vec{B}$, for
  any bidder $i$, we have $P_i \ge 0$.
\item[Pareto-optimality (PO):] We must have (i) $\sum_i X_i = 1$, {\em
    i.e.}, the good is completely sold; and (ii) If $X_i > 0$ and $v_j >
  v_i$, then $P_j = B_j$, {\em i.e.}, if a bidder gets non-zero quantity
  then all bidders with higher valuations have exhausted their budgets. This property holds ex-post (regardless of randomization).
\end{description}

In the case of $m$ indivisible units, the only difference is in the (PO)
condition. This gets modified as: $\sum_i X_i = m$; further, if $X_i > 0$
and $v_j > v_i$, then $v_i > B_j - P_j$. In both cases, this corresponds to
the fact that no pair of agents can improve their utility by trading.

\medskip The main focus of this paper is to understand the behavior of the
adaptive clinching auction, which is described in Section~\ref{sec:clinch}.
For a more detailed description (especially for the indivisible units case),
please see Dobzinski {\em et al}~\cite{nisan}. Their main result is the
following:

\begin{theorem}[Dobzinski {\em et al}~\cite{nisan}]
\label{thm:nisan}
The adaptive clinching auction satisfies (VP), (NPT), and (PO) with private
budgets and valuations. When the budgets are public knowledge, the auction
also satisfies (IC), and it is the unique auction satisfying (VP), (NPT), (PO),
and (IC) ex-post. Furthermore, there is no auction satisfying these four
properties ex-post when the budgets are private.
\end{theorem}

Our main goal is to show several structural results about this auction,
which will culminate in showing that there is indeed a {\em randomized}
mechanism  that is (VP) and (IC) {\em in expectation}, and also satisfies (NPT) and
Pareto-optimality ex-post (regardless of the outcome of randomization), even with private budgets.

\subsection{Budget Monotonicity and its Consequences}  
\label{sec:res}
We will show that the adaptive clincing auction with one infinitely
divisible good satsifies  Budget Monotonicity, which states that
a bidder cannot gain by reporting a lower budget.
\begin{define}
A deterministic auction is  Budget Monotone if the following
conditions hold for every bidder $i$ regardless of $v_{-i},
B_{-i}$. For reported budget $B_i \in [0,\beta_i]$, where $\beta_i$ is the
true budget:
\begin{enumerate}
\item The bidder always maximizes utility by reporting
  $v_i = \eta_i$, where $\eta_i$ is the true valuation.
\item When $v_i = \eta_i$, the utility of the bidder is {\em monotonically
    non-decreasing} in $B_i \in [0,\beta_i]$.
\end{enumerate}
\end{define}

The more interesting condition in the above definition is the second one (the first one following from~\cite{nisan}).
Though Budget Monotonicity is an intuitive property, there is no guarantee that it
is satisfied even by reasonable auctions. In fact, quite surprisingly, it
does not always hold for the adaptive clinching auction!

\begin{theorem}
\label{thm:eg}
  In the case of $m = 4$ indivisible units of a good and $n = 3$ bidders, the
  adaptive clinching auction described in~\cite{nisan} does not satisfy Budget Monotonicity.
\end{theorem}
\begin{proof}
Consider $n = 3$ bidders with the following $(B_i, v_i)$ values: $(B_1,
  v_1) = (6,3)$, $(B_2, v_2) = (5,3)$, and $(B_3, v_3) = (4,3)$, and
  suppose these are the true budgets and valuations. It is easy to show
  that bidders $1$ and $2$ clinch one unit each at price $2$. Bidder $3$ obtains zero utility since she can only clinch when price is $3$.
  However, if bidder $3$ reports $(3,3)$, she clinches one unit at price
  $17/6$ and obtains strictly positive utility. Therefore, the auction is
  not monotone. The details are easy to fill in using the description in~\cite{nisan}.
\end{proof}

In sharp contrast, our main result is to show that the adaptive clinching
auction indeed satisfies the  Budget Monotonicity property when there is one
infinitely divisible good (which is the limiting case of $m$ indivisible
goods). In particular, our main theorem is the following:

\begin{theorem}[{\bf Budget Monotonicity Theorem}]
\label{thm:main}
The adaptive clinching auction satisfies  Budget Monotonicity
for one infinitely divisible good.
\end{theorem}

The key intuitive difference between the divisible and indivisible cases is that in the former case, there is a nice characterization of bidders receiving non-zero allocations as those with highest remaining budgets (refer Lemma~\ref{lem:clinch}).  Budget Monotonicity is equivalent to saying that a
bidder cannot gain by under-reporting her budget, {\em i.e.}, reporting $B_i < \beta_i$. We now show a simple way to remove the
incentive to report $B_i > \beta_i$.

\medskip\noindent{\bf Randomized Extraction:} We run the deterministic adaptive clinching auction as in~\cite{nisan}. The allocation remains the same. However, the price extraction scheme is randomized as follows. If a bidder reports budget $B_i$ and is supposed to pay $P_i \in \left[0,B_i\right]$ according to the deterministic mechanism, then with probability $P_i/B_i$, we extract her reported budget $B_i$, and with probability $\left(1 - P_i/B_i\right)$, we charge her {\em zero} price.  Note that this randomization can be applied to any deterministic auction where $X_i > 0$ implies $P_i > 0$; the adaptive clinching auction~\cite{nisan} does satisfy this property. It is now easy to show:

\begin{theorem}
\label{lem:up}
  For the case of one infinitely divisible good, the randomized adaptive clinching
  auction satisfies (NPT) always, (VP) and (IC) in expectation, and is Pareto-optimal ex-post.
\end{theorem}
\begin{proof}
  Clearly, (NPT) is always satisfied. The expected payment after randomization is precisely $P_i$, which
  preserves (VP).   To see (PO), observe that
  if $P_j = B_j$ before the randomization, the same is true after the
  randomization. To show (IC), note that the auction satisfies Budget Monotonicity
  by Theorem~\ref{thm:main}, so that for any bidder $i$, we have $v_i = \eta_i$ and $B_i \ge \beta_i$. Furthermore, if bidder $i$ reports a budget $B _i > \beta$ and recevies nonzero allocation, then the deterministic adaptive clinching auction charges her a price $P_i > 0$. The randomized auction extracts $B_i$ w.p.
  $P_i/B_i > 0$ and in this scenario, the utility of the bidder is $-\infty$.
  Therefore, the bidder will not report $B_i > \beta_i$. 
\end{proof}

If the allocations generated by the auction for the infinitely divisible
case are treated as probabilities of allocation instead (similar to~\cite{abrams,borgs}), the same auction
works for the case of indivisible units of the good.

\begin{corollary}
\label{cor:manygoods}
There is a randomized auction satisfying (NPT) always, (VP) and (IC) in
expectation, and is Pareto-optimal ex-post for $m$ indivisible units of the good.
\end{corollary}
\begin{proof} Run the randomized adaptive clinching auction assuming one infinitely divisible good, with the valuations scaled
  up by factor of $m$. Modify the allocation step as follows.  Suppose the auction should allocate $x_i \in [0,1]$ to
  bidder $i$. Choose bidder $i$ with probability $x_i$ and
  allocate all $m$ units to this bidder. The resulting auction always satisfies (NPT). Also note that the expected utility of
  a bidder is the same as the utility in the indivisible auction; further,
  the auction is (IC) by Theorem~\ref{lem:up}. To show (PO), note that
  the items are completely allocated by the auction. Next, since the
  infinitely divisible auction satisfies (PO), if the $m$ units are
  allocated to bidder $i$, this bidder must have had $X_i > 0$ in the
  infinitely divisible auction, so that for all $j$ with $v_j > v_i$, we
  must have $P_j = B_j$, so that $v_i > B_j - P_j = 0$. Therefore, the
  auction satisfies (PO) regardless of the outcome of randomization.
\end{proof}

\section{The Adaptive Clinching Auction: Infinitely Divisible Case}
\label{sec:clinch}
We now describe the adaptive clinching auction in~\cite{nisan,ausubel} in
the context of one infinitely divisible good, and show in the next section
that it satisfies Budget Monotonicity (Theorem~\ref{thm:main}).

Intuitively, the auction is an ascending price auction. As the price per
unit quantity is raised, bidders become inactive because the price has
exceeded their valuation. For the remaining (active) bidders, the {\em
  demand} is the amount they are willing to buy given their remaining
budget and the current price. Similarly, the {\em supply} is the amount of
item remaining. For an active bidder, when the supply exceeds the total
demand of the other bidders, this bidder {\em clinches} the difference at
the current price. Since the price is increased continuously and the good
is infinitely divisible, the auction defines a differential process.

We describe the clinching auction as a differential process indexed by time
$t$, where the price charged per unit quantity increases as time
progresses, and the auction continuously allocates (part of) the item and extracts
budget. We note that the traditional method is to describe it as a process
indexed by the price; however, indexing by time lends itself to an easier
analysis.  After describing the auction, we present some new observations
that characterize its behavior; these will be useful in later sections.

Formally, let $p(t)$ denote the price per unit quantity at time $t \ge 0$.
For bidder $i$, let $x_i(t)$ denote the quantity of the item allocated so
far to $i$, let $P_i(t)$ denote the price extracted so far from $i$, and let $b_i(t)$
denote the {\em effective} budget of the bidder (defined later). Let $S(t)
= 1 - \sum_i x_i(t)$ denote the {\em supply} of item left with the
auctioneer.  Initially, $p(0) = P_i(0) = x_i(0) = 0$, $b_i(0) = B_i$, and
$S(0) = 1$. In this section, we will denote the derivative of function $g(t)$
w.r.t. $t$ as $g'(t)$.

Denote the {\em demand} of the bidder as $D_i(t) =
\frac{b_i(t)}{p_i(t)}$. If $p(t) < v_i$, this represents the amount of
the item bidder $i$ is willing to buy at price $p(t)$. Let $D_{-i}(t) =
\sum_{j \neq i} D_j(t)$,  {\em i.e.}, the total demand excluding bidder $i$.

\medskip\noindent{\bf Invariants.} Denote the stopping time of the auction by $f$.
The adaptive clinching auction is defined by the following invariants for all $t <
f$:

\begin{description}
\item[Supply Invariant:] For all bidders $i$, we have $S(t) \le D_{-i}(t) = \sum_{j \neq i}
  D_j(t)$. 
\item[Clinching Invariant:] $x'_i(t) > 0$ iff both $p(t) < v_i$ (the bidder is
  {\em active}) and $S(t) = D_{-i}(t)$.  
\item[Budget Invariant:] If $p(t) < v_i$, $b_i(t) = B_i - P_i(t)$, the true
  residual budget of the bidder. If $p(t) > v_i$, $b_i(t) = 0$. When $p(t)
  = v_i$, $b_i(t) \in [0,B_i - P_i(t)]$, and though the demand $D_i(t)$ is well-defined, it will not correspond to any ``real'' demand, since the bidder will drop out of the auction. 
\end{description}

We note that for all $t < f$, we have $S'(t) = - \sum_i x'_i(t)$.
Furthermore, we also have $x'_i(t) = - \frac{b'_i(t)}{p(t)}$, since the
bidder is being charged price $p(t)$ per unit quantity. The only exception
to these conditions is at time $f$ when the auction makes some one-shot
allocations; we will define $S(f) = \lim_{t \rightarrow f} S(t)$.

\subsection{Auction} 
In view of the above invariants, we partition the bidders into the following groups.

\begin{define}
  Define  {\em active} bidders as $A(t) = \{j | v_j > p(t) \mbox{ and }
  b_i(t) > 0 \}$; {\em exiting} bidders as $E(t) = \{j | v_j = p(t)
  \mbox{ and } b_i(t) > 0\}$; and {\em clinching bidders} as $C(t) =
  \{j | j \in A(t) \mbox{ and } S(t) = D_{-i}(t)\}$.
\end{define}

\begin{figure}[htbp]
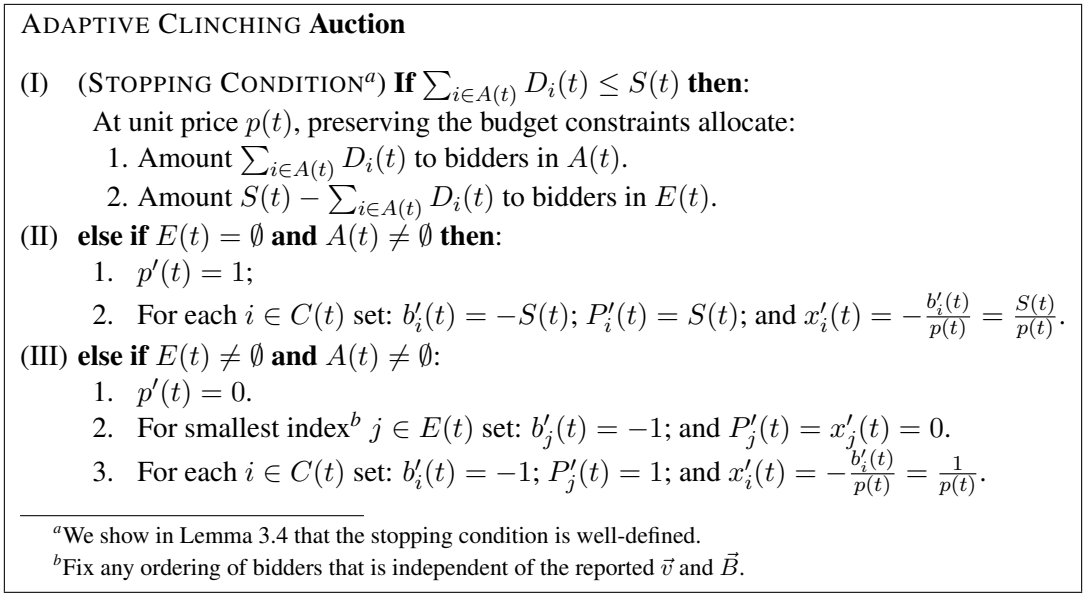

\centerline{\framebox{
\begin{minipage}{4.5in}
{\bf {\sc Adaptive Clinching} Auction}
\begin{tabbing}
  \= (I)\ \ \  \ \= ({\sc Stopping Condition\footnote{We show in Lemma~\ref{lem:bud} that the stopping condition is well-defined.}}) {\bf If} $\sum_{i \in A(t)} D_i(t) \leq S(t)$ {\bf then}: \\
  \> \> \ \ \=At unit price $p(t)$, preserving the budget constraints allocate:  \\
  \>\> \> \ \ 1.  \= Amount $\sum_{i \in A(t)} D_i(t)$ to bidders in $A(t)$. \\  
  \>\>\> \ \ 2. \> Amount $S(t) - \sum_{i \in A(t)} D_i(t)$ to bidders in $E(t)$.  \\ 
  \>(II)\> {\bf else if} $E(t) = \emptyset$ {\bf and} $A(t) \neq \emptyset$ {\bf then}:  \\ 
  \>\>\>  1. \> $p'(t) = 1$; \\
  \>\>\> 2. \> For each $i \in C(t)$ set:  $b'_i(t) = -S(t)$;  $P'_i(t) =
  S(t)$; and $x'_i(t) = - \frac{b'_i(t)}{p(t)} = \frac{S(t)}{p(t)}$. \\ 
  \> (III)\> {\bf else if} $E(t) \neq \emptyset$ {\bf and} $A(t) \neq \emptyset$: \\ 
  \>\> \>1. \> $p'(t) = 0$. \\
  \>\> \>2. \> For smallest index\footnote{Fix any ordering of bidders that is independent of the reported $\vec{v}$ and $\vec{B}$.}  $j \in E(t)$ set: $b'_j(t) = -1$; and $P'_j(t) =
  x'_j(t) = 0$. \\ 
  \>\> \>3. \> For each $i \in C(t)$ set:  $b'_i(t) = -1$; $P'_j(t) = 1$; and
  $x'_i(t) = - \frac{b'_i(t)}{p(t)} = \frac{1}{p(t)}$. 
\end{tabbing}
\end{minipage}
}}
\caption{\label{fig:clinch} The adaptive clinching auction for one
  infinitely divisible good.} 
\end{figure}

The adaptive clinching auction is now simple to describe, and is described
in Figure~\ref{fig:clinch}. We specify it in
terms of the derivatives of the budget, allocation, and prices.

The total allocation $X_i$ and the total price $P_i$ can easily be derived
from the description of the auction; we omit the details.  Note that
a bidder {\em clinches} items only when the allocation is made and he is in $A(t)$. Though the bidder
may get some items in Step (I) when he is in $E(t)$, we do not consider this {\em clinching},
since the bidder gets utility zero from these items (assuming she reports
the true valuation). 

The key difference between the way we have described the auction and that in~\cite{nisan} is in Step (III). Here, we have chosen to gradually reduce the budgets of the bidders in $E(t)$, while if the auction were indexed by price, this step would lead to one-shot allocations. Our method makes the supply $S(t)$ and the effective budgets $b_i(t)$ continuous functions. The equivalent formulation of Step (III) in terms of price follows from maintaining the Supply Invariant and stopping condition of the auction (see also~\cite{nisan}), and is presented below.

\begin{lemma}
\label{lem:adjust}
If $t < f$ and $i \in A(t)$, suppose $\sum_{j \in A(t), j \neq i} D_j(t) < S(t)$, then bidder $i$ clinches $S(t) - \sum_{j \in A(t), j \neq i} D_j(t)$ quantity at price $p(t)$ in Step (III). When $t = f$ and $A(f) \neq \emptyset$, bidder $i \in A(f)$ clinches a quantity at price $p(f)$ that exhausts her  remaining budget in Step (I).
\end{lemma}

The above lemma will be critically used in the proof of Budget Monotonicity later. The next theorem simply re-states the positive result in Theorem~\ref{thm:nisan}.

\begin{theorem}[Dobzinski {\em et al}~\cite{nisan}]
\label{thm:val}
The adaptive clinching auction satisfies (NPT), (VP), (PO). Furthermore,
for reported budget $B_i \le \beta_i$ where $\beta_i$ is the true budget,
the bidder always maximizes utility by reporting the true valuation, $v_i =
\eta_i$.
\end{theorem}

\subsection{Properties}
We will now show some properties of this auction that will be useful later.

\begin{define}
\label{def:bud}
Define $b_{\max}(t) = \max_{i \in A(t)} b_i(t)$. Recall that $f$ as the
stopping time of the auction.
\end{define}

We first show that the auction satisfies the invariants.  The last two
invariants are easy to check: Whenever a bidder $i$ clinches the item at
price $p(t) < v_i$, we have $b'_i(t) = - p(t) x'_i(t) = -P'_i(t)$, so that
$b_i(t) = B_i - P_i(t)$. Further, note that if $i \in E(t)$, the effective
budget $b_i(t)$ of this bidder reduces, so if the price increases beyond
$v_i$, the effective budget must be identically $0$.  Therefore, the budget
invariant holds. The clinching invariant holds trivially by the description
of the auction. 

The next result shows the supply invariant, and characterizes the set of bidders that are clinching at any point in time, and a consequent stopping condition based on these bidders.

\begin{lemma}
\label{lem:sup}
\label{cor_bmax_clinch}
\label{lem:clinch}
The following hold for the adaptive clinching auction: (1)  If $C(t) \neq \emptyset$, then $C(t) = \{j \in A(t) | b_j(t) = b_{\max}(t)\}$; (2) the supply invariant holds for all $t < f$; and (3) when a bidder $i$ drops from $C(t)$, the auction stops at that time.
\end{lemma} 
\begin{proof}
  First note that for $t < f$, the functions $S(t)$ and $D_{-i}(t)$ for any
  $i \in A(t)$ are continuous. At $t = 0$, the former is smaller than the
  latter.  If for all $t < f$, $S(t) < D_{-i}(t)$, then there is nothing to
  prove. Suppose $t < f$ be the first time instant when $S(t)$ becomes
  equal to $D_{-i}(t)$. If $i \notin A(t)$, then $\sum_{j \in A(t)} D_j(t)
  \leq D_{-i}(t) = S(t)$, the auction necessarily stops at time $t$, that
  is, $t = f$, a contradiction. Thus, bidder $i \in A(t)$. Furthermore, $i$
  has the largest budget in the set $A(t)$, since for all $j \neq i$, we
  must have had $D_{-j}(t) \ge D_{-i}(t) = S(t)$ by the definition of time
  $t$. Therefore, $b_i(t) = b_{\max}(t)$ Since $D_{-i}(t) = S(t)$, we have
  $i \in C(t)$. For all $t' < t$, the set $C(t)$ is empty by the clinching
  invariant. We now show that for all subsequent $t$, as long as $i$ has
  not dropped out, we have $S'(t) = D'_{-i}(t)$; $b_i(t) = b_{\max}(t)$;
  and when $i$ drops out of the auction, the auction stops. This will show all parts of the
  lemma.

  First note that if $C(t)$ is non-empty, it necessarily has the bidders
  with highest budget in $A(t)$. Therefore, if $i \in C(t)$, then $b_i(t)$
  is necessarily the same as $b_{\max}(t)$. To show $i$ clinches continuously, we will
  show $S'(t) = D'_{-i}(t)$ when $i \in C(t)$, so that for all $t$ as long
  as bidder $i$ has not dropped out, $S(t) = D_{-i}(t)$, and hence bidder $i$
  clinches continuously until she drops out.

  Suppose $i \in C(t)$ at some point $t$ and $p(t) < v_i$. We therefore
  must have $b_i(t) = b_{\max}(t)$ and $S(t) = D_{-i}(t)$. Note that $S(t)$
  decreases at the rate of precisely $- \sum_{j \in C(t)} x'_j(t) = c
  \frac{b'_{\max}(t)}{p(t)}$ if there are $c$ clinching bidders at time $t$.

  There are two cases depending on whether $E(t) = \emptyset$ or not. If $E(t) =
  \emptyset$, then for all $j \in C(t)$:
$$D'_j(t) =  \frac{p(t) b'_j(t) - b_{\max}(t) p'(t)}{p(t)^2} = - \frac{S(t)}{p(t)} - \frac{b_{\max}(t)}{p(t)^2}$$
Similarly, for $j \in A(t) \setminus C(t)$, we have:
$$ D'_j(t) = - \frac{b_j(t)}{p(t)^2}$$
Note that $S'(t) =   c \frac{b'_{\max}(t)}{p(t)}  = - c \frac{S(t)}{p(t)}$. Since $i \in C(t)$, we have:
$$ \frac{d}{dt}\left(D_{-i}(t)\right) = - (c-1) \frac{S(t)}{p(t)} -
\frac{D_{-i}(t)}{p(t)} = - (c-1) \frac{S(t)}{p(t)} - \frac{S(t)}{p(t)} = -
c \frac{S(t)}{p(t)} = S'(t)$$ 

Suppose now that $E(t) \neq \emptyset$ and $A(t) \neq \emptyset$. Note that $b'_k(t)
= -1$ for some $k \in E(t)$. We have: $D'_j(t) = 0$ for $j \notin C(t) \cup
\{k\}$. For $j \in C(t) \cup \{k\}$ we have:
\begin{eqnarray*}
D'_j(t)  & = &  \frac{p(t) b'_j(t) - b_{\max}(t) p'(t)}{p(t)^2} = - \frac{1}{p(t)} \\
\Rightarrow \ \  \frac{d}{dt}\left(D_{-i}(t)\right) & = & -\frac{c}{p(t)} = - c \frac{b'_{\max}(t)}{p(t)} = S'(t)
\end{eqnarray*} 

Therefore, $S'(t) = D'_{-i}(t)$ if $i \in C(t)$, which shows bidder $i$
clinches continuously unless $b_i(t) = 0$ or $p(t) = v_i$. In both cases,
$\sum_{j \in A(t)} D_j(t) \le S(t)$ so that Step (I) kicks in and the
auction stops.
\end{proof}

The above characterization of $C(t)$ holds only for infinitely divisible goods, and is the key reason Budget Monotonicity holds in this case and not in the case of indivisible units. Also note that the auction could stop even if no bidders drop from $C(t)$, but instead, some other set of bidders drop out; therefore, part (3) in the above lemma is a sufficient but not necessary condition for stopping.

The above lemma establishes that for $t < f$, the function $b_{\max}(t)$ is continuous; further, the set $C(t)$, if non-empty, is composed of active bidders $i \in A(t)$ with $b_i(t) = b_{\max}(t)$.  We now show that the stopping condition (Step (I)) is well-defined, and relate the prices charged to the stopping condition.

\begin{lemma}
\label{lem:bud}
When the auction stops, the bidders in $A(f) \cup E(f)$ have sufficient budget to clinch the quantity
$S(f)$ at price $p(f)$.
\end{lemma}
\begin{proof}
  By defintion of the stopping time, for time $t$ approaching $f$ from below, we have $\sum_{i
    \in A(t)} D_i(t) \geq S(f)$. Note that $A(f) \cup E(f) \subseteq
  A(t)$. The lemma follows.
\end{proof}

\begin{lemma}
\label{cor_Bi}
If $i \notin A(t) \cup E(t)$, then $b_i(t) = 0$. If $i \in A(t) \setminus
C(t)$, then $b_i(t) = B_i$. Furthermore, if $p(f) < v_i$, then $P_i =
B_i$ and $b_i(f) = 0$.
\end{lemma}

The following lemma shows that the amount of randomness we need to add is small. In particular, we need to randomize the
price charged to {\em at most one} bidder.

\begin{lemma}
\label{lem:onlyone}
The allocations in Step (I) can be done in a fashion so that when the auction stops,  there is at
most one bidder $i$ with allocation $X_i > 0$ and price $P_i < B_i$.
\end{lemma}
\begin{proof}
  At time $f$   the only bidders
  who can have $P_i \in (0, B_i)$ are bidders in $E(f)$.  As $t$ approaches $f$ from below, suppose some
  bidder in $C(t)$ dropped out causing the auction to stop. At time $t$, the supply invariant
  holds from the perspective of this bidder who drops out, so that if this
  bidder is given lowest priority in allocating the remaining supply, the
  supply exhausts the budget of all bidders except this bidder.  If no
  clinching bidder drops out or if $C(t) = \emptyset$, then all $i \in E(f)$ had
  $X_i = 0$, so that the budget can be extracted sequentially from bidders
  in $E(f)$. This satisfies the lemma.
\end{proof}

\section{The Budget Monotonicity Theorem}
\label{sec_proof}
In this section, we will provide a proof sketch of Theorem~\ref{thm:main} for a canonical special case of budgets and valuations; the entire proof is complicated with many cases, and is presented in Appendix~\ref{app:monotone}. The observation that
the bidder will always report $v_i = \eta_i$ follows from~\cite{nisan},
re-stated in Theorem~\ref{thm:val}.  We will now show that when $v_i =
\eta_i$, a bidder does not gain utility by reporting budget $B_i <
\beta_i$, where $\beta_i$ is the true budget. This will complete the proof
of Theorem~\ref{thm:main}, and hence all the results in
Section~\ref{sec:res}. 

\subsection{Notation}
 We fix a specific bidder, say {\em Alice}, and show
monotonicity of her utility with reported budget. We will use sub-script
$*$ to denote quantities for this bidder.  Let $b_{*}(t)$ and $v_{*}$
respectively denote her effective budget at time $t$, and her valuation.
Let $P_*(t)$ represent the price extracted from her so far.

For convenience, we will use $t_-$ to denote the limit as $x$ approaches
time $t$ from below. Since price increases continuously with time, we can
easily replace $p(t_-)$ by $p(t)$ in any algebraic expression. However, if
 $t$ is the first time instant when the price becomes equal to the valuation of some bidder, then $\left\{i \ | \ v_i >
  p(t)\right\} \subset \left\{i \ | \ v_i > p(t_-)\right\}$, and so on.

\medskip
\noindent {\bf Formula for Utility.} For times $t' \le t'' \leq f$, let
$u(t',t'')$ denote the utility gained by Alice as time increased from $t'$
to $t''$. In the computation of utility, we can ignore the contribution
from allocation made in Fig.~\ref{fig:clinch} when Alice is in $E(f)$,
since the allocation is obtained at a price equal to her valuation.  If $x_{*}(t)$ is
the fraction of the item clinched by Alice until time $t$, then, for $t'' < f$ in
 Steps (II) and (III) of the auction:

\begin{equation}
\label{eq_utility_db}
u(t',t'') = \int_{t'}^{t''} (v_{*} - p(t)) \frac{d}{dt}\left(x_{*}(t)
\right)dt = \int_{t'}^{t''} - \frac{(v_{*} - p(t))}{p(t)} \frac{d}{dt}\left(b_{*}(t)
\right)dt
\end{equation}
Moreover, when $t'' = f$, the formula gets modified by the one-shot allocation in Step (I):
$$u(t',f) = \frac{(v_* - p(f))}{p(f)} b_*(f_-) + \int_{t'}^{f_-}
\frac{(v_{*} - p(t))}{p(t)} \frac{d}{dt}\left(b_{*}(t) 
\right)dt $$ Define $u$ as the the total utility gained by Alice from the
auction.

\medskip 
\noindent {\bf Two Auctions.} We let Alice increase her reported budget by
an amount $\Delta > 0$, the budgets and valuations of other bidders and
Alice's valuation and true budget remaining the same. Suppose her original reported budget is $B^0_*
\le \beta_*$, and her new reported budget is $B^1_* = B^0_* + \Delta \le
\beta_*$.  Denote the former auction (with Alice's reported budget being $B^0_*$) by
{\sc Low} and the latter auction by {\sc High}. We will use superscripts
$0$ and $1$ to denote quantities in these two auctions respectively. Note
that for $i \neq$ Alice, we have $B_i^1 = B_i^0$.

\medskip
We will show the following theorem (proved in Appendix~\ref{app:monotone}), which will imply the proof of
Theorem~\ref{thm:main}. This will also imply all results in Section~\ref{sec:res}.

\begin{theorem}
\label{th_MAIN}
$u^0 \le u^1$, {\em i.e.}, Alice's utility from auction {\sc High}
is at least her utility from auction {\sc Low}.
\end{theorem}

\subsection{Proof Sketch}
The proof of Theorem~\ref{th_MAIN} is very technical and is hence relegated to Appendix~\ref{app:monotone}. We outline the basic argument for a special case where the valuations are sufficiently large so that Alice clinches for a finite amount of time in both the auctions. The following definition describes the times at  which Alice starts and stops clinching, and the time at which the clinching set becomes nonempty. 

\begin{define}
\label{def:start_time}
Let $y^0$ (resp. $y^1$) denote the first time instant at which some bidder
enters the clinching set in auction {\sc Low} (resp. {\sc High}), that is,
$C^0(t)$ (resp. $C^1(t)$) becomes nonempty. Similarly, define $q^0$ (resp.
$q^1$) to be the first time when Alice enters the clinching set in
auction {\sc Low} (resp. {\sc High}).
\end{define}

\noindent {\bf Simplifying Assumptions.} The valuations of the bidders are sufficiently large so that Alice clinches in both auctions, {\em i.e.},  $y^0 \leq q^0 < f^0$ and $y^1 \leq q^1 < f^1$. Alice has the minimum valuation amongst all the bidders, so that no other bidder drops out before Alice, and by Lemma~\ref{lem:clinch}, the auction stops when Alice drops out, so that $f^0 = f^1 = f$ and $p(f) = v_*$. Moreover, the highest budgeted bidder (say bidder $1$) has larger budget than Alice in both the auctions {\sc Low} and {\sc High}, that is, $B_1 > B_*^1 = B_*^0 + \Delta$.

We now track the two auctions simultaneously as time increases from zero. First note that since Step (III) in Figure~\ref{fig:clinch} is never executed, and price increases at rate $1$ in Step (II), as long as both auctions run, the prices in the two auctions are coupled as time progresses, and further, the set $A(t)$ is the set of all bidders. Therefore, we can use price and time interchangeably. Since Alice cannot gain any utility after dropping out, we have the following for the utilities of Alice in either auction: $u^0 = u^0(q^0,f_-)$ and $u^1 = u^1(q^1,f_-)$.  The main ingredient in the proof is to show the following relation, which implies that though {\sc Low} starts clinching before {\sc High}, Alice starts clinching in {\sc High} earlier than when she starts clinching in {\sc Low}:
\begin{lemma}
$y^0 < y^1 \leq q^1 \leq q^0 < f $.
\end{lemma}
\begin{proof}   
First note that since bidder $1$ has the maximum initial budgets in both the auctions, by Lemma~\ref{lem:clinch}, it must be the case that bidder $1$ starts clinching in {\sc Low}  at time $y^0$, and in {\sc High} at time $y^1$. Beyond time $y^0$ in {\sc Low}, the quantity $b^0_{\max}(t)$ decreases with time at rate equal to the supply, $S^0(t)$. Since $S^0(t) \le 1$, the rate of decrease is at most $1$, which easily implies: 
\begin{equation}
\label{eq:blah}
b^0_{\max}(t) \geq b^0_{\max}(y^0) + (y^0 - t)
\end{equation}

Next note that $b^0_{\max}(y^0) = B_1 > B_*^1 = B_*^0 + \Delta$.   Combining the above relations, we have $b^0_{\max}(y^0 + \Delta) > B_*^0$ and hence, by the Clinching Invariant, Alice is not clinching at time $y^0 + \Delta$, so that by definition, $q^0 > y^0 + \Delta$. It is straightforward to see that $y^1 = y^0+ \Delta$, since the total of the budgets of bidders $2,3,\ldots,n$ differs  by exactly $\Delta$, and hence the Clinching Invariant kicks in  $\Delta$ time later. Therefore,  $q^0 > y^0+ \Delta = y^1$. 

We will next show that $q^1 \in [y^1, q^0]$. Note that both auctions are clinching beyond $y^1$. Using Equation (\ref{eq:blah}) at $t=y^1$, we must have $B_1 = b^1_{\max}(y^1) = b^0_{\max}(y^0) \leq b^0_{\max}(y^1) + \Delta$. Next, we use the observation (Lemma~\ref{lem:sup}) that the clinching sets $C(t)$ in the two auctions are related to the value $b_{\max}(t)$, which decreases at rate $S(t)$. The clinching set  if non-empty is precisely $\{i | b_i(t) = b_{\max}(t)\}$. Therefore the auction with the larger $b_{\max}(t)$ has a smaller clinching set.  Using this, we show that in the auction with larger $b_{\max}(t)$, this value decreases at a faster rate. Specifically, if $b_{\max}^1(t) \geq b^0_{\max}(t)$ for $t \ge y^1$, then for all bidders $i$, $b_i^1(t) \geq b^0_i(t)$, and hence, the demands are larger in auction {\sc High}. By the Clinching Invariant, this implies the supply $S(t)$ is larger in auction {\sc High}, and hence, the rate of decrease of $b_{\max}(t)$ is larger.
$$b_{\max}^1(t) \geq b^0_{\max}(t) \qquad \Rightarrow \qquad \frac{d}{dt}\left( b_{\max}^1(t) \right) \le \frac{d}{dt}\left( b^0_{\max}(t) \right) \leq 0\qquad \forall \ t \ge y^1$$
 Since  $b^1_{\max}(y^1)  \leq b^0_{\max}(y^1) + \Delta$, we must have for all $t < f$: $b^1_{\max}(t) \leq b^0_{\max}(t) + \Delta$. Specifically, $b^1_{\max}(q^0) \leq b^0_{\max}(q^0) + \Delta = B_*^0 + \Delta = B_*^1$. Thus, $q^1 \leq q^0$, which completes the proof. 
 \end{proof}
 
 This shows Alice is clinching in both auctions beyond $q^0$, which helps us relate her utilities. Observe by Eq. (\ref{eq_utility_db}) that the utilities are related to the rate of decrease of $b_{\max}(t)$. Using a similar argument to the above:
$$ b^1_{\max}(t) \ge b^0_{\max}(t) \qquad  \Leftrightarrow \qquad \frac{d}{dt}b^1_{\max}(t) \le \frac{d}{dt}b^0_{\max}(t) \le 0 \qquad \forall t \ge q^0$$
We consider two cases. First, if $b^1_{\max}(q^0) \geq b^0_{\max}(q^0)$, then we have $\frac{d}{dt}\left( b_{\max}^1(t) \right) \le \frac{d}{dt}\left( b^0_{\max}(t) \right)$, and $b^1_{\max}(t) \ge b^0_{\max}(t)$ for all $t \ge q^0$. Applying Equation~\ref{eq_utility_db}, we have:
$$ u^1 \geq \int_{q^0}^{f_-} -\frac{\left(v_* - p(t)\right)}{p(t)}\frac{d}{dt}\left(b^1_{\max}(t)\right) \geq \int_{q^0}^{f_-} -\frac{\left(v_* - p(t)\right)}{p(t)}\frac{d}{dt}\left(b^0_{\max}(t)\right) = u^0 $$   
If $b^1_{\max}(q^0) < b^0_{\max}(q^0)$,  then we need to take into account the utility gained by Alice in {\sc High} during the time interval $q^1 \leq t \leq q^0$ to complete the proof; details are in Appendix~\ref{app:monotone}.                    

The reason the general case is complicated is that we need to take care of two tricky issues: (1) One of the auctions can stop due to bidders dropping out. We need to account for this event in several of the proofs. (2) If $q^0 = f^0$, then Alice obtains only a one-shot allocation in Step (I) of the auction.  In this case, we have an explicit formula for the utility of Alice in auction {\sc Low}, and we essentially argue that the auction {\sc High} derives at least that much utility at price $p(q^0)$. This shows Theorem~\ref{th_MAIN} already holds in this case.

\section{Bayesian Setting and Size Constraints}
\label{app:size}
We now consider the case where the bidders have a public size constraint on
the amount of item they want to buy in addition to a private valuation and
private budget. In this setting, we consider optimizing social welfare and revenue.
Performing this optimization in the adversarial setting is difficult: For
instance, it is no longer true that the optimal single price auction yields
a constant factor approximation to the revenue~\cite{Goldberg04}; and
further, the adaptive clinching auction could in fact yield zero revenue if
the sum of the size constraints is less than one.  Instead, we consider the Bayesian setting and assume
the auctioneer maintains independent discrete distributions on the possible
valuations and budgets for each bidder, and is interested in designing a
(randomized) mechanism for optimizing the expected revenue (respectively
social welfare). 

Variants of this model have been considered before~\cite{myerson,vohra,laffont,maskin,shuchi,timr}, and the optimal solution can indeed be encoded as an (exponential size) linear program. The key challenge now becomes designing  polynomial time computable mechanisms. It is well-known~\cite{myerson,vohra} that the optimal mechanism has a simple structure related to the VCG mechanism in the case of {\em  i.i.d.} distributions and no size constraints. It is unlikely that such a structure holds in the general setting, and we instead consider designing approximately optimal mechanisms. The budget constraints however make a poly-time relaxation of the problem non-linear. However, if we only encode that utility decreases for under-reporting budgets, the program becomes linear;  we again use randomization to prevent over-reporting budgets.  Using this, we show a poly-size linear program relaxation with a rounding scheme that yields a $5.83$ approximation to the optimal Bayesian IC mechanism  when the type space is discrete; this mechanism is  implementable in dominant strategies. This rounding technique may be applicable in other related scenarios.

We consider mechanisms that are implementable in dominant strategies,  meaning  that the bidders may or may not be aware of these densities, and are simply interested in maximizing utility, where the utility is in expectation over the randomness introduced by the mechanism. Note that
while the auctioneer optimizes over the densities, the bidders simply
optimize over the randomness introduced by the mechanism and not over the
densities themselves.

\medskip \noindent {\bf Problem Statement.} Formally, there are $n$ bidders, and one unit of an infinitely
divisible good. Bidder $i$ has private valuation $v_i$ per unit for the
good, which can take on values $0 = s_0 \le s_1 < s_2 < \cdots < s_K$.
Further, the bidder has a private budget $b_i$ and the budgets can take on
values $0 = \beta_0 \le \beta_1 < \beta_2 < \cdots < \beta_M$. The bidder
is interested in acquiring a maximum $\kappa_i$ amount of the item, and
this value is public knowledge.  The auctioneer maintains an independent discrete distribution over possible (valuation, budget) pairs of bidder $i$. Let $f_{ikm} = \Pr[v_i = s_k \mbox{
  and } b_i = \beta_m]$. The type space of a bidder is discrete and we will be interested in mechanisms that can be computed in time polynomial in the input size, {\em i.e.}, in the quantities $n, K, M$.
  
  For reported bid vector $\vec{v}, \vec{b}$, the
auctioneer computes allocations $x_i(\vec{v},
\vec{b})$ and prices $P_i(\vec{v}, \vec{b})$ so that the resulting auction
satisfies (VP), (NPT), and (IC). Since we will use randomness to force a bidder to not over-report the budget, these properties will be in expectation over the randomness
introduced by the auction. In addition, the allocation satisfies the size
constraints: $x_i(\vec{v}, \vec{b}) \le \kappa_i$. Subject to these
constraints, the auctioneer is interested in maximizing either: (1)
Revenue, $\E[\sum_i P_i(\vec{v},\vec{b})]$, where the expectation is over
the distributions $f_{ikm}$ from which the $\vec{v}, \vec{b}$ are drawn; or
(2) Social welfare, $\E[\sum_i v_i x_i(\vec{v},\vec{b})]$, where the
expectation is as before.

\medskip
\noindent {\bf Preventing Over-reporting Budgets.} We will again construct a deterministic auction that assumes bidders cannot over-report budgets, and introduce randomness so that (VP), (NPT), and (IC) hold in expectation over the randomness introduced. Since we can no longer guarantee that non-zero allocation implies non-zero price, the randomization is slightly different from the  Randomized Extraction Scheme: Suppose the deterministic auction makes non-zero allocation $X_i$ and charges price $P_i$ (which could be zero). Then, extract price $P_i$ initially; then, with some probability $\delta \in (0,1)$, extract additional price $B_i - P_i$, and independently with probability $\delta$, give the bidder amount $B_i - P_i$. This makes the expected price charged exactly $P_i$; however, since there is a non-zero probability of extracting $B_i$, this prevents the bidder from reporting $B_i$ larger than the true budget constraint as this would make the expected utility $-\infty$. This also preserves (VP), (IC), and (NPT) in expectation over the randomness. We term this the {\sc Threat}; the key difference from the randomized extraction scheme is that the {\sc Threat} preserves (NPT) only in expectation.

\medskip
Therefore, we can now restrict the bidder to not over-report her budget constraint and consider deterministic mechanisms that are ex-post (IC) subject to this restriction. The optimal mechanism maximizing expected revenue (resp. socially welfare) and satisfying the above
constraints can be encoded as an LP of exponential size. We note that the optimal solution to the restricted mechanism design problem is also an upper bound on the revenue (resp. social welfare) that can be obtained for mechanisms that do not explicitly restrict bidders to not over-report budgets.  

We now show a $5.83$-approximation mechanism whose computation time is polynomial in the input size, {\em i.e.}, in the quantities $n,K,M$.  In the rest of the discussion, we focus on revenue maximization; the results for social welfare are almost identical to derive.

\subsection{Linear Programming Formulation}
Let $x_{ikm} = \E_{\vec{v}_{-i}, \vec{b}_{-i}}[x_i(s_k,b_m,\vec{v}_{-i},
\vec{b}_{-i})]$, and let $P_{ikm} =\E_{\vec{v}_{-i},
  \vec{b}_{-i}}[P_i(s_k,b_m,\vec{v}_{-i}, \vec{b}_{-i})]$. These are respectively the expected allocation and price charged for bidder $i$ if she reports $(s_k,b_m)$, where the expectation is over the values revealed by the other bidders according to the distributions $f_{j}, j \neq i$.  Consider the
following linear program, essentially due to Myerson~\cite{myerson}.

\[ \mbox{Maximize } \ \ \ \sum_{i,k,m} f_{ikm} P_{ikm} \]
\[ \begin{array}{rcll}
\sum_{i,k,m} f_{ikm} x_{ikm} & \le & 1  \\
s_{k} x_{ikm} - P_{ikm}  & \ge & s_{k} x_{ilt} - P_{ilt}  & \forall i, k,
l,\text{ and } \forall t < m\\ 
s_{k} x_{ikm} - P_{ikm}  & \ge & 0  & \forall i, k,  m\\
P_{ikm} & \in & [0,b_m] & \forall i, k, m \\
x_{ikm} & \in & [0,\kappa_i] & \forall i, k, m\\
\end{array} \]

We note that the above program linearizes the utility constraint by only
encoding that the utility of under-reporting the budget is at most that of reporting the true budget. (The auction will finally introduce
the {\sc Threat} to prevent lying in the other direction.) Attempting to
encode lying in {\em both directions} in the above program (as done by Pai
and Vohra~\cite{vohra}) makes it non-linear.

\begin{lemma}
\label{lem:my3}
The truthful deterministic auction maximizing revenue is feasible for the
above constraints. Therefore, the LP value is an upper bound on the
expected revenue
\end{lemma}
\begin{proof}
  Consider the optimal truthful auction. Since (VP), (NPT), (IC), and the
  budget and size constraints hold ex-post, they hold in expectation over
  any independent densities $f_{ikm}$. The first, third, fourth, and fifth
  constraint simply encode feasibility and (VP). The second constraint
  maintains (IC), since otherwise, a bidder gains in utility by lying
  either on her valuation, or downward on her budget.  Therefore, the
  optimal deterministic auction is feasible for the constraints of the LP. 
\end{proof}

\subsection{The Auction}
The first step is to solve the linear program, which can be done in time
polynomial in $n, K, M$. The linear program does not directly yield a
feasible auction. Consider any realization $\vec{v}, \vec{b}$ of the bids
where bidder $i$ reports $(s_{k_i}, b_{m_i})$. The linear program indicates
that $x_{ik_im_i}$ amount of the item should be alloted to agent $i$.
However, it can happen that $\sum_i x_{ik_im_i} > 1$ in this realization.
Note that the LP only enforces the constraint $\sum_i x_{ik_im_i} \le 1$
{\em in expectation} over the reported $\vec{v}, \vec{b}$, where the
expectation is over the densities $f_{ikm}$; enforcing it for all $\vec{v},
\vec{b}$ would need exponentially many variables and constraints. We note
that in the absence of budget constraints or in the presence of {\em
  i.i.d.} distributions with no size constraints, the optimal
solution to the above LP has a very simple structure~\cite{myerson,vohra}
related to the VCG mechanism. It is not clear if the structure holds in the
presence of budget and size constraints.

We now convert the LP solution into a feasible mechanism by losing a factor $5.83$ in
the worst case. For any $\alpha > 1$ consider the following mechanism:

\begin{enumerate}
\item Scale down all variables in the LP by a factor of $\alpha$. Let
  $\hat{x}_{ikm} = \frac{x_{ikm}}{\alpha}$ and $\hat{P}_{ikm} =
  \frac{P_{ikm}}{\alpha}$.
\item Consider the bidders in a fixed but arbitrary order $1,2,3,\ldots,
  n$. Suppose bidder $i$ reports $(s_{k_i}, b_{m_i})$:
\item Let $z_i = 1 - \sum_{j < i} \tilde{x}_{j}$. If $z_i<\frac{1}{\alpha}$
  then allocate 0 and charge 0 to bidder $i$; if $z_i \ge \frac{1}{\alpha}$
  then:
\begin{enumerate}
\item Allocate $\tilde{x}_i = \hat{x}_{ik_im_i} $ units to bidder $i$.
\item Charge price $\tilde{P}_i = \hat{P}_{ik_im_i}$.
\end{enumerate}
\end{enumerate}

\begin{theorem} Let $y_{ikm}$ and $q_{ikm}$ denote the expected allocation
  and price in the above auction when agent $i$ bids $(s_k, b_m)$, where
  the expectation is over the densities $f_{jkm}$ for $j \neq i$. We have
  the following:
\begin{enumerate}
\item $\tilde{x}_i \le x_{ikm} \le \kappa_i$; $\tilde{P}_{i} \le P_{ikm}
  \le b_m$; and $\sum_i \tilde{x}_i \le 1$. 
\item $y_{ikm} \ge x_{ikm} \frac{\alpha-2}{\alpha(\alpha-1)}$ and $q_{ikm}
  \ge P_{ikm} \frac{\alpha-2}{\alpha(\alpha-1)}$.
\item The auction satisfies (VP), (IC), and (NPT) ex-post assuming bidders cannot over-report budgets.
\end{enumerate}
Therefore, when the above auction is randomized via the {\sc Threat},
it is a $5.83$ approximation to the optimal expected revenue, implementable in dominant strategies, and satisfies (VP), (IC), and (NPT) in expectation over the randomness introduced by the {\sc Threat}.
\end{theorem}
\begin{proof}
  The first part is straightforward from the description of the auction. To
  see the third part, note that the allocation and price for bidder $i$ are
  either both zero, or are $\hat{x}_{ikm} = \frac{x_{ikm}}{\alpha}$ and
  $\hat{P}_{ikm} = \frac{P_{ikm}}{\alpha}$. The choice between zero and the
  latter cannot be affected by $i$ by changing its bid. Since
  $\hat{x}_{ikm}, \hat{P}_{ikm}$ satisfy the constraints of the LP, this
  shows the third part of the theorem.

  We now show the second part.  From the perspective of the auctioneer, the
  auction has a random outcome that depends on the densities $f_{ikm}$. Let
  $W_i$ denote the random variable which is $1$ if $z_i \ge
  \frac{1}{\alpha}$, and $0$ otherwise. Let $w_i = \E[W_i]$.  We first show
  that $w_i \ge \frac{\alpha-2}{\alpha-1}$.  We have $\E_{\vec{v}_{-i},
    \vec{b}_{-i}}[\sum_{j \neq i} \hat{x}_{j k_j m_j}] \le
  \frac{1}{\alpha}$. By Markov's inequality, $\Pr_{\vec{v}_{-i},
    \vec{b}_{-i}}\left[ \sum_{j \neq i} \hat{x}_{j k_j m_j} \ge 1-
    \frac{1}{\alpha}\right] \le \frac{1}{\alpha-1}$. If this does not
  happen, we must have the event $W_i =1$. Therefore, $w_i \ge
  \frac{\alpha-2}{\alpha-1}$.

  Note that in the event $W_i = 1$, the allocation to $i$ is $\hat{x}_{ikm}
  \le \frac{1}{\alpha} \le z_i$. Next note that $W_i$ is independent of the
  bid reported by $i$. Therefore, $y_{ikm} = \hat{x}_{ikm} w_i \ge x_{ikm}
  \frac{\alpha-2}{\alpha(\alpha-1)}$ and $q_{ikm} = \hat{p}_{ikm} w_i \ge
  p_{ikm} \frac{\alpha-2}{\alpha(\alpha-1)}$, which shows the second part.

  Note that the revenue generated is $\sum_{ikm} f_{ikm} q_{ikm}$, which is
  a $\frac{\alpha(\alpha-1)}{\alpha-2}$ approximation by the above theorem.
  If we set $\alpha = 2 + \sqrt{2}$, we obtain a
  $\frac{\alpha(\alpha-1)}{\alpha-2} = 3 + 2\sqrt{2} = 5.83$ approximation
  to the optimal expected revenue.
\end{proof}

\appendix

\section{Proof of Budget Monotonicity: Theorem~\ref{th_MAIN}}
\label{app:monotone}
This section is devoted to the proof of Theorem~\ref{th_MAIN}. We note
that by Theorem~\ref{thm:val}, the reported valuations are always the
truth, meaning that $v_i = \eta_i$ for all bidders $i$.   

Recall that bidder Alice increases her budget by a quantity $\Delta$ in auction {\sc High} as compared to auction {\sc Low}. Also recall that the subscript $*$ is used to denote quantities for Alice, and the superscripts $1, 0$  to denote quantities in auctions {\sc High} and {\sc Low} respectively.  We will define the following starting and stopping times.

\begin{define}
\label{def:start}
Let $y^0$ (resp. $y^1$) denote the first time instant at which some bidder
enters the clinching set in auction {\sc Low} (resp. {\sc High}), that is
$C^0(t)$ (resp. $C^1(t)$) becomes nonempty. Similarly, define $q^0$ (resp.
$q^1$) to be the first time instant when Alice enters the clinching set in
auction {\sc Low} (resp. {\sc High}).  If the required event does not
happen, define these as $f^0$ (resp. $f^1$).
\end{define}

Recall that we will use $t_-$ to denote the limit as $x$ approaches
time $t$ from below. Since price increases continuously with time, we can
easily replace $p(t_-)$ by $p(t)$ in any algebraic expression. However, if
 $t$ is the first time instant when the price becomes equal to the valuation of some bidder, then $\left\{i \ | \ v_i >  p(t)\right\} \subset \left\{i \ | \ v_i > p(t_-)\right\}$, and so on.

\subsection{Assumptions } 
\label{sec:assume}
We now show that the theorem is straightforward if some assumptions do not
hold. First, note that if $p^0(q^0) = v_*$, Alice receives zero utility in
{\sc Low}, and  Theorem~\ref{th_MAIN} is trivially true. Thus, we must have:

\medskip
\begin{figure}[htbp]
\centerline{
\framebox{
\begin{minipage}{6in}
\begin{assumption}
\label{as_p(q0)<v*}
Alice receives non-zero utility in auction {\sc Low}. In other words, $u^0 > 0$ and $p^0(q^0) < v_{*}$.
\end{assumption} 
\end{minipage}}
}
\end{figure}

Using this assumption, we show that the prices in the two auctions are
coupled. Let $p^0(t)$ and $p^1(t)$ denote the prices at time $t$ in {\sc
  Low}, {\sc High} respectively.

\begin{claim}
\label{cl_p(t)_same}
For all $t \leq \min(f^0,f^1)$, $p^0(t) = p^1(t)$
\end{claim}  
\begin{proof}
  At the beginning, $p^0(0) = p^1(0) = 0$. Simultaneously follow auction
  {\sc Low} and {\sc High} as time increases from zero. When the price is
  not equal to the valuation of any bidder, both $p^0(t)$ and $p^1(t)$ are
  increasing at unit rate. When the price $p$ hits the valuation of some
  bidder(s), two cases may occur. If the set $\left\{i \ | \ v_i =
    p\right\}$ has nonempty intersection with $C^0(t_-)$ (resp.
  $C^1(t_-)$), then auction {\sc Low} (resp. {\sc High}) necessarily stops
  at that time $t = \min(f^0,f^1)$. Otherwise, if none of the bidders with
  valuation equal to $p$ belonged to $C^0(t_-) \cup C^1(t_-)$, then
  price remains equal to $p$ in both the auctions for exactly $\sum_{i \ :
    \ v_i = p} B_i$ amount of time. Note that in the later case, Alice
  cannot have a valuation equal to $p$, else she receives zero utility in
  both the auctions and Theorem~\ref{th_MAIN} is trivially true.
\end{proof}

From now on, we will use $p(t)$ to denote both $p^1(t)$ and $p^0(t)$. A
direct consequence of the above proof is the following, whose proof is
simple and omitted. Note that $E(t)$ are coupled since the auctions do not stop (so that all bidders in $E(t)$ could not have been clinching), and Step (III) reduces the budgets of these bidders in a fixed order.

\begin{corollary}
\label{cl_active_same}
For all $t < \min(f^0,f^1)$, $A^0(t) = A^1(t)$. Further, $E^0(t) = E^1(t)$.
\end{corollary}

We will now show another assumption whose violation easily implies 
Theorem~\ref{th_MAIN}.

\medskip
\begin{figure}[htbp]
\centerline{
\framebox{
\begin{minipage}{6.0in}
\begin{assumption}
\label{as_stop_time}
Auction {\sc High} stops at a time that is strictly greater than the price
at which Alice starts to clinch in auction {\sc Low}, that is $f^1 >
q^0$.
\end{assumption}
\end{minipage}}
}
\end{figure}

\begin{claim}
\label{cl_stop_time}
If Assumption~\ref{as_stop_time} is violated, Theorem~\ref{th_MAIN} is true.
\end{claim}
\begin{proof}
  Suppose $f^1 \leq q^0$. Clearly, $q^0 \leq f^0$. From
  Assumption~\ref{as_p(q0)<v*}, $p(f^1) \leq p(q^0) < v_*$. From the
  Lemma~\ref{cor_Bi}, $P_*^1 = B_*^1$. In auction {\sc High}, Alice receives at least
  $\frac{B^1_*}{p(f^1)}$ fraction of the item at an average unit price
  that is at most $p(f^1)$. That is, $$u^1 \geq \frac{v_* -
    p(f^1)}{p(f^1)}B^1_*$$ However, in auction {\sc Low}, she can
  receive at most $\frac{B^0_*}{p(q^0)}$ fraction of the item at an
  average unit price that is at least $p(q^0)$. That is, $$u^0 \leq
  \frac{v_* - p(q^0)}{p(q^0)}B^0_*$$ Since $p(f^1) \leq p(q^0)$ and
  $B_*^1 > B_*^0$, we get $u^1 \geq u^0$. This implies
  Theorem~\ref{th_MAIN}.
\end{proof}

We will use Assumptions~\ref{as_p(q0)<v*} and~\ref{as_stop_time} several times throughout the rest of Appendix~\ref{app:monotone}.

\subsection{The Canonical Case: Alice Enters Set $C(t)$ in Auction {\sc Low}, that is $q^0 < f^0$}
\label{app:times}
The argument consists of two stages. First we relate the times at which Alice starts to
clinch in either auction, in particular, we show that Alice starts clinching in {\sc High} no later than in {\sc Low}. This statement is critically used in the next stage of our proof, where we compare the utilities gained by Alice in the two auctions, and show that her utility from {\sc Low} is at most her utility from {\sc High}.

\begin{lemma}[Structure Lemma]
\label{lm_main}
The starting and stopping times in {\sc High} and {\sc Low} are related as:  $$y^0 \leq y^1 \leq q^1 \le q^0 < \min(f^0,f^1)$$
\end{lemma}

Since $q^0 < f^0$, Assumption~\ref{as_stop_time} immediately implies the last inequality. Most important part of the above lemma is the claim that $q^1 \leq q^0$, {\em i.e.}, Alice joins the clinching set no later in {\sc High} than in {\sc Low}.    

\subsubsection{Proof of the Structure Lemma}
\label{sec:struct}
Let bidder $1$ have the largest budget among all active bidders excluding Alice at the time when auction {\sc Low} starts clinching. We first present a high-level idea of the proof. At any point in time, the set of clinching bidders, if non-empty, is the set of active bidders $i$ with $b_i(t) = b_{\max}(t)$; furthermore, once the auction starts clinching, $b_{\max}(t)$ decreases continuously.  We therefore relate the evolution of $b_{\max}(t)$ in the two auctions, and show the time $t$ at which $b_{\max}(t) = B^0_*$ in auction {\sc Low}  is at least the time $t$ at which $b_{\max}(t) = B^1_*$ in auction {\sc High}.   We use the following observations about the curves $b^0_{\max}(t)$ and $b^1_{\max}(t)$ in auctions {\sc Low} and {\sc High} respectively:

\begin{enumerate}
\item  The curves have downward slope at most $1$, and are parallel.
\item  Auction {\sc High} starts clinching at most $\gamma$ time after {\sc Low} starts clinching, where $\gamma = \min(B_1,B^*_1) - B^0_*$. In particular, {\sc High} starts clinching before time $q^0$; and
\item If $b^0_{\max}(t) \le b^1_{\max}(t)$, then $b^1_{\max}(t)$ decreases at a faster rate once both auctions are clinching. 
\end{enumerate}

Using these observations, the proof is simple geometry with two cases depending on whether $B_1 \le B^1_*$ or  $B_1 \ge B^1_*$, {\em i.e}, whether or not Alice has the highest budget in auction {\sc High}. We now present the proof in the following sequence of claims.

\begin{claim}
\label{cl_very_begin}
$y^0 \leq y^1$. Furthermore, for all $t \leq y^0$, we have the following: If bidder $i$ is not Alice, then $b^0_i(t) = b^1_i(t)$. In
particular, $b^0_i(t) = b^1_i(t) = B_i$ when $p(t) < v_i$. For Alice, $b^0_*(t) = B^0_* < b^1_*(t) = B^1_*$.
\end{claim}
\begin{proof}
As time $t$ increases gradually from $t=0$,  as long as no bidder is clinching in either auction ({\em i.e.}, $C^0(t) = C^1(t) = \emptyset$), the current budget $b_i(t)$ of every active bidder $i$ equals her original budget $B_i$. Furthermore, Step (III) reduces the budgets of exiting bidders in a fixed order. We conclude that the current budget of every bidder other than Alice remains the same across the two auctions, and the current budget of Alice is greater in auction {\sc High} than in {\sc Low}. Thus, from the perspective of any bidder (including Alice), the total demand of the other bidders is no less in auction {\sc High} than in auction {\sc Low}. In particular, it implies {\sc Low} starts clinching no later than {\sc High}, that is, $y^0 \leq y^1$. 
\end{proof}

\begin{lemma}
\label{lem_struct_case3}
If Alice is the first bidder to join the clinching set $C^0(t)$ in auction {\sc Low}, {\em i.e.}, if $y^0 = q^0$, then the Structure Lemma holds.
\end{lemma}
\begin{proof}
Suppose $y^0 = q^0$. From Alice's perspective, in both the auctions, total demand of other bidders is exactly equal to the initial supply at time $t = y^0$ (Claim~\ref{cl_very_begin}). Thus, Alice joins the clinching set in {\sc High} at the same time instant as in {\sc Low}. Hence $y^0 = y^1 = q^1 = q^0$ and the Structure Lemma holds. 
\end{proof}

For the rest of Section~\ref{sec:struct}, we assume that Alice is {\em not} the first bidder to start clinching in {\sc Low}, so that $y^0 < q^0$. Therefore bidder $1$ with $B_1 > B^0_*$ starts clinching in {\sc Low} at time $y^0$.  So far, we have the following inequalities:

\begin{equation}
\label{eq_times_1} 
y^0 < q^0 < \min\left(f^0,f^1\right) \qquad \mbox{and} \qquad y^0 \leq y^1 \qquad \mbox{and} \qquad B_1 > B^0_*
\end{equation}  

The above implies  $b^0_{\max}(y^0) = B_1 = B_*^0 + \delta$ for some $\delta > 0$. Note that the active sets $A(t)$ in the two auctions are identical at any point in time (Corollary~\ref{cl_active_same}). Applying Lemma~\ref{lem:clinch} and Assumption~\ref{as_p(q0)<v*}, we have that both Alice and bidder $1$ are active during the time interval $y^0 \leq t \leq q^0$. Furthermore, for all $t \in \left[y^0,q^0\right]$, bidder $1$ belongs to the clinching set in {\sc Low}, and has the maximum budget amongst all the active bidders, that is, $b^0_{\max}(t) = b^0_{1}(t)$.

\begin{claim}
\label{lm_rate_bmax}
In auctions {\sc Low} and {\sc High}, $b_{\max}(t)$ decreases at a rate at most one, {\em i.e.},  $ \frac{d}{dt} \left(b_{\max}(t)\right) \in [-1, 0]$.
\end{claim} 
\begin{proof}
If the clinching set $C(t)$ is empty, then $\frac{d}{dt} \left(b_{\max}(t)\right) = 0$. Otherwise, by Lemma~\ref{lem:clinch}, either $\frac{d}{dt} \left(b_{\max}(t)\right) = -1$ in Step (III), or $\frac{d}{dt} \left(b_{\max}(t)\right) = -S(t)$ in Step (II). Since $S(t) \leq 1$, the claim follows.
\end{proof}

\begin{claim}
\label{cl_gap_y0q0}
Recall $\delta = B_1 - B^0_*$. In auction {\sc Low}, Alice starts clinching at least $\delta$ later than the time instant at which bidder $1$ starts clinching, {\em i.e.},  $q^0 \ge y^0 + \delta$.
\end{claim}
\begin{proof}
As time increases beyond $y^0$, by Lemma~\ref{lem:clinch}, Alice starts clinching in {\sc Low} when $b^0_{\max}(t)$ becomes equal to Alice's reported budget $B_*^0$. Since $ b^0_{\max}(y^0) = B_1 =  B_*^0 +  \delta $, and since $b^0_{\max}(t)$ decreases at a rate at most one, we have the claim.
\end{proof}

\begin{claim}
\label{cl_geometry}
Suppose bidder $1$ has higher initial budget than Alice in  {\sc High} {\em i.e.}, $B_1 > B_*^1$,  then  we have $y^1 = y^0 + \Delta < q^0$.
\end{claim}
\begin{proof}
Since $B_*^0 + \delta = B_1 > B_*^1 = B_*^0 + \Delta$, we have $\Delta < \delta = B_1 - B^0_*$. Thus, the inequality $y^0 + \Delta < q^0$ follows from Claim~\ref{cl_gap_y0q0}. Applying the Clinching Invariant and Claim~\ref{cl_very_begin} in {\sc High}, we have at  $t = y^0$:
$$p(t) = p(t) S^0(t) = b^0_{-1}(t) = b^1_{-1}(t) - \Delta$$

As time increases beyond $y^0$, as long as there is no clinching in {\sc High}, either the LHS
increases at rate $1$ in Step (II) or the RHS decreases at rate one in Step
(III). In either case, at time $t = y^0 + \Delta$, we must have $p(t) =
b^1_{-1}(t)$, which implies $y^1 = y^0 + \Delta$. 
\end{proof}

\begin{claim}
\label{cl_flip}
Suppose bidder $1$ has higher initial budget than Alice in auction {\sc High} {\em i.e.}, $B_1 > B_*^1$. If $b^1_{\max}(t) \geq b^0_{\max}(t)$ at some time instant $t \in \left[y^1,q^0\right)$, then $\frac{d}{dt}\left(b^1_{\max}(t)\right) \leq
\frac{d}{dt}\left(b^0_{\max}(t)\right) \leq 0$.
\end{claim}
\begin{proof}
  Note that clinching set is nonempty in both auctions in this time range. The active and exiting sets, $A(t)$ and $E(t)$, in the two auctions are coupled (Corollary~\ref{cl_active_same}). If the existing set $E(t)$ is nonempty, then $b^0_{\max}(t)$ and $b^1_{\max}(t)$ are each decreasing at rate one in Step (III), and the claim is true. For rest of the proof, assume $E(t)$ is empty.

We first show that compared to auction {\sc Low}, every active bidder has larger remaining budget in auction {\sc High}, {\em i.e.}, $b^1_i(t) \ge b^0_i(t)$. Since $b^1_{\max}(t) \geq b^0_{\max}(t)$, the statement is clearly true for all bidders who are clinching either in {\sc High} or in {\sc Low}. For all other active bidders, the current budget $b_i(t)$ equals the reported budget $B_i$. Since Alice reports a higher budget in {\sc High} and every other bidder reports the same budget in the two auctions, the statement is valid even for active bidders who are not clinching in {\em both} auctions. 
  
Since Alice reports a lower budget than bidder $1$ in {\sc High}, bidder $1$ is clinching in {\sc High} the time range $\left[y^1,q^0\right)$. Considering the Supply Invariant from the perspective of bidder $1$, we conclude that
\begin{equation}
\label{eq_simplify}
S^1(t) - S^0(t)  =  \sum_{i \neq 1} \frac{b^1_i(t)}{p(t)} - \sum_{i \neq
  1} \frac{b^0_i(t)}{p(t)} = \sum_{i \ : \ i \in A(t), i \neq 1} \frac{b^1_i(t) - b^0_i(t)}{p(t)} \geq 0    
\end{equation}
The above holds since the exiting set $E(t)$  is empty and since all active bidders have larger remaining budget $b_i(t)$ in {\sc High}. The claim
follows immediately from Step (II) of Figure~\ref{fig:clinch}.
\end{proof}

\begin{figure}
\centerline{\includegraphics[width=4.00in]{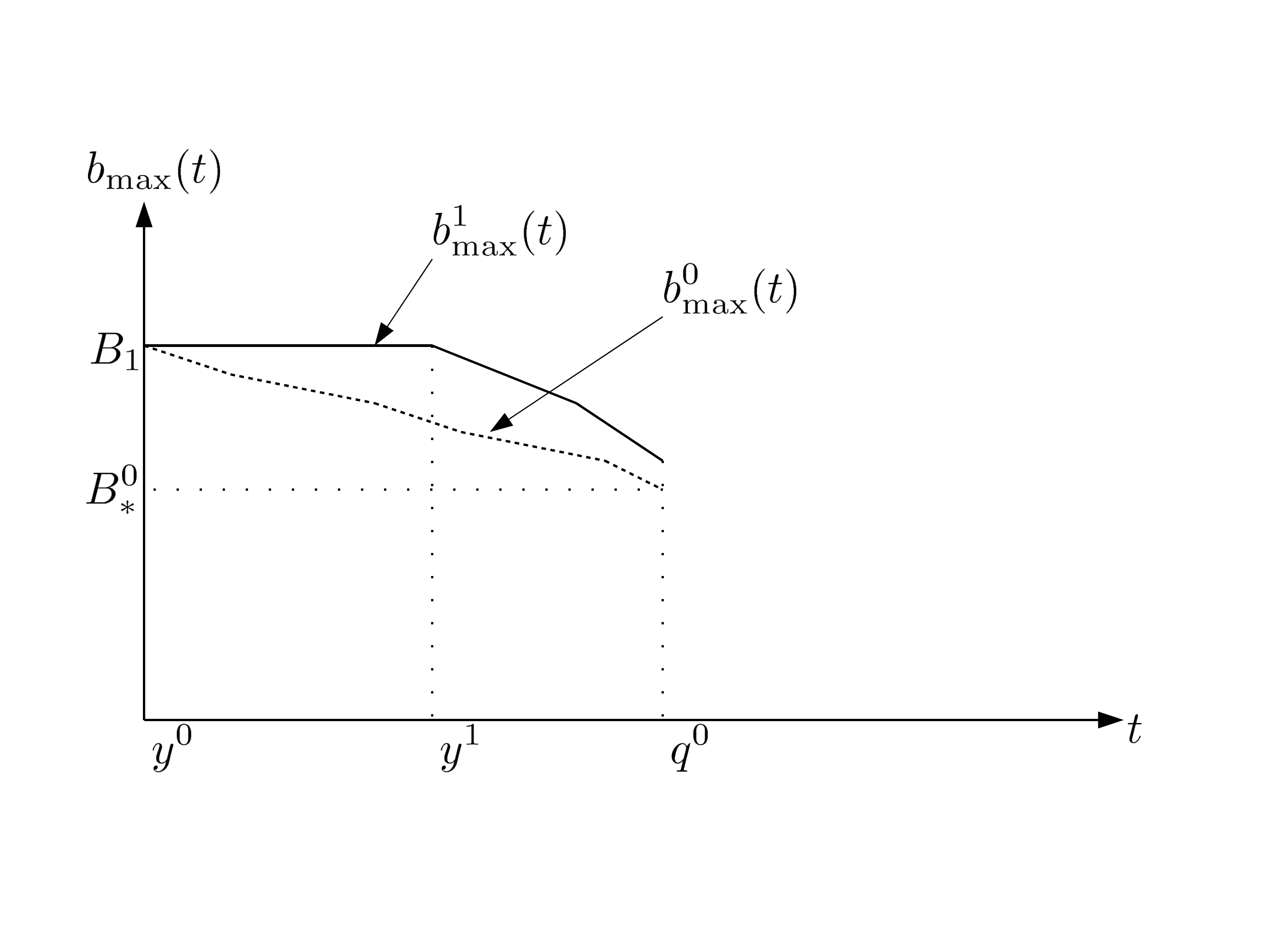}}
\vspace{-0.5in}
\caption{Proof of the Structure Lemma for the case $B_1 > B^1_*$. Note that {\sc High} starts clinching at most $\Delta = B^1_* - B^0_*$  time later, and beyond this point, for any $t$, $b^1_{\max}(t)$ decreases at least as fast as $b^0_{\max}(t)$.}
\label{fig:struct}
\end{figure}

\begin{lemma}[Lemma~\ref{lm_main}]
The starting and stopping times in {\sc High} and {\sc Low} are related as:  $$y^0 \leq y^1 \leq q^1 \le q^0 < \min(f^0,f^1)$$
\end{lemma}
\begin{proof} All we need to show is $q^1 \le q^0$. We split the proof into cases depending on whether Alice has the highest budget in {\sc High} or not.

\paragraph{Case 1. $B_*^1 \geq B_1$:}  At time $t = y^0$, in both auctions the current budget of every active bidder equals her initial budget (Claim~\ref{cl_very_begin}). Since Alice reports a higher budget than bidder $1$ in {\sc High}, Alice has the highest budget amongst all the active bidders in {\sc High} at time $t = y^0$. By Assumption~\ref{as_p(q0)<v*} and Claim~\ref{cl_gap_y0q0}, Alice is active during the time interval $\left[y^0, q^0\right] \supseteq \left[y^0,y^0+\delta\right]$. As time increases beyond $y^0$, as long as Alice is active, no other bidder can start clinching before Alice in auction {\sc High} (Lemma~\ref{lem:clinch}). Considering the Clinching Invariant for auction {\sc Low}, at time $t = y^0$,
\begin{eqnarray*}
p(t) = p(t) S^0(t) = b_{-1}^0(t) = b_{-\mbox{Alice}}^1(t) - \delta
\end{eqnarray*}
The last equality follows from Claim~\ref{cl_very_begin}. In auction {\sc High}, in the time range $[y^0, y^0+\delta]$, either the
LHS increases at rate $1$ in Step (II) or the RHS decreases at rate $1$ in Step (III), so that Alice must start clinching at $y^1 = q^1 = y^0+\delta$. Combining this with Claim~\ref{cl_gap_y0q0}, we have the proof.

\paragraph{Case 2. $B_1 > B^1_*$:} Since Alice's reported budget in {\sc High} is less than that of bidder $1$, bidder $1$ has the maximum budget amongst active bidders in {\sc High} at time $t = y^0$ (Claim~\ref{cl_very_begin}). By Lemma~\ref{lem:clinch}, at every time instant $t \in \left[y^0,q^0\right]$, we have $b^1_{\max}(t) = b^1_1(t)$. Therefore, during the interval $\left[y^0,q^0\right]$, in {\em both} auctions, $b_{\max}(t)$ is  equal to the current budget of bidder $1$.  We simultaneously track $b_{\max}(t)$ of the two auctions in this time range (see Figure~\ref{fig:struct}). At time $t = y^0$, both $b^0_{\max}(t)$ and $b^1_{\max}(t)$ are equal to $B_1$ (initial budget of bidder $1$). Bidder $1$ starts clinching in {\sc Low} at the same time instant. Thus, $b^0_{\max}(t)$ starts  decreasing continuously as $t$ increases beyond $y^0$. However, $b^1_{\max}(t)$ decreases below $B_1$ only after $t$ goes past the value $y^1$ (note that $y^1 \geq y^0$ by Claim~\ref{cl_very_begin}). Claim~\ref{cl_geometry} shows that $y^1$ and $y^0$ differs by exactly $\Delta = B^1_* - B^0_*$ amount. In particular, since $y^1 \leq q^0$, $b^1_{\max}(t)$ starts decreasing before Alice enters the clinching set in {\sc Low}. Applying Claim~\ref{lm_rate_bmax}, at time time $t = y^1$, the vertical distance between the curves $b^0_{\max}(t)$ and $b^1_{\max}(t)$ is no more than $\Delta$.  Claim~\ref{cl_flip} implies that in the time range $y^1 \leq t < q^0$, whenever the curve $b^1_{\max}(t)$ lies above $b^0_{\max}(t)$, the former reduces at a larger rate. We thus have $b^1_{\max}(q^0) \leq b^0_{\max}(q^0) + \Delta$. Since Alice starts clinching in {\sc Low} at time $q^0$, $b^0_{\max}(t)$ equals her reported budget $B^0_*$ at that time instant. In other words, $b^1_{\max}(q^0) \leq B^0_* + \Delta = B_*^1$, and Alice must have joined the clinching set in {\sc High} no later than $q^0$. This completes the proof.
\end{proof}

Figure~\ref{fig:struct} illustrates the geometric intuition behind the above proof in the case $B_1 > B^1_*$.

\subsubsection{Relating the Utilities in Auctions {\sc Low} and {\sc High}}
By Lemma~\ref{lem:sup}, Alice clinches in {\sc Low} (resp. {\sc High})
throughout the time interval $q^0 \leq t < f^0$ (resp. $q^1 \leq t < f^1$).
During the next phase of our proof, we simultaneously track the two
auctions as time increases from $q^0$ to $\min(f^0,f^1)$ and show that
$b_{\max}(t)$ of one of the auctions dominates the other. This helps us
compare the utilities gained by Alice during this phase, including the
utilities from one-shot allocations at the stopping times.

Define $f_{\min} = \min(f^0,f^1)$. Note that the Structure Lemma implies
$\max(y^0,y^1) < f_{\min}$. In particular, by Lemma~\ref{lem:clinch}, Alice
clinches in auction {\sc High} (resp. {\sc Low}) during the time interval
$q^1 \leq t < f_{\min}$ (resp. $q^0 \leq t < f_{\min}$). We will need the
following two claims. The proofs are similar to that of Claim~\ref{cl_flip}. By the Structure Lemma, Alice is clinching in {\em both} auctions when $q^0 \leq t < f_{\min}$, and we only need to replace bidder $1$ by Alice in Equation~\ref{eq_simplify}.

\begin{claim}
\label{cl_rate_bmax_3}
\label{cl_lower_budget2}
If $b^1_{\max}(t) \geq b^0_{\max}(t)$ at some time $q^0 \leq t <
f_{\min}$, then $\frac{d}{dt}\left(b^1_{\max}(t)\right) \leq
\frac{d}{dt}\left(b^0_{\max}(t)\right) \leq 0$. Therefore, if
$b^1_{\max}(q^0) < b^0_{\max}(q^0)$, then for all $t \in [q^0,f_{\min})$,
$b^1_{\max}(t) \leq b^0_{\max}(t)$.
\end{claim}

\begin{claim}
\label{cl_rate_bmax_2}
\label{cl_greater_budget}
If $b^1_{\max}(t) \leq b^0_{\max}(t)$ at some time $q^0 \leq t < f_{\min}$,
then $\frac{d}{dt}\left(b^0_{\max}(t)\right) \leq \frac{d}{dt}
\left(b^1_{\max}(t)\right) \leq 0$. Therefore, if $b^1_{\max}(q^0) \geq
b^0_{\max}(q^0)$, then for all $t \in [q^0,f_{\min})$, we have
$b^1_{\max}(t) \geq b^0_{\max}(t)$.
\end{claim}

\subsubsection{Case 1: $b^1_{\max}(q^0) \ge b^0_{\max}(q^0)$} 
We will now prove Theorem~\ref{th_MAIN} in two cases. We will first prove Theorem~\ref{th_MAIN} under the assumption that
$b^1_{\max}(q^0) \ge b^0_{\max}(q^0)$. 

\begin{lemma}
\label{lem:less}
If $b^0_{\max}(q^0) \leq b^1_{\max}(q^0)$, then $u^0 \le u^1$. 
\end{lemma}
\begin{proof}
  We first show $f^0 \leq f^1$. Suppose $ f^0 > f^1 $. If some bidder $i$
  other than Alice is clinching in {\sc High} just before time $f^1$, then
  $B_i \geq b_i^1(f^1_-) = b_{\max}^1(f^1_-) \geq b_{\max}^0(f^1_-)$. Thus,
  $i \in C^0(f^1_-)$. Also note that Alice $\in C^0(f^1_-) \cap
  C^1(f^1_-)$. Thus, $C^1(f^1_-) \subseteq C^0(f^1_-)$. If auction {\sc
    High} stops at time $f^1$ because some bidder in $C^0(f^1_-)$ drops
  out, then {\sc Low} will also stop at $f^1$, a contradiction. Thus,
  assume none of the bidders with valuation equal to $p(f^1)$ is in the set
  $C^0(f^1_-)$. All those bidders will retain their initial budgets in both
  the auctions till time $f^1_-$. Therefore the {\em excess-demand} (that
  is, $\sum_{k \ : \ v_k > p(t)} D_k(t) - S(t)$) will reduce by the same
  quantity in both the auctions at time $t = f^1$. Now, by the clinching
  invariant, the difference between the excess demands between auctions
  {\sc High} and {\sc Low} at time $f^1_-$ is precisely
  $\frac{b^1_{\max}(f^1_-) - b^0_{\max}(f^0_-)}{p(f^1)}$, so that since
  $b_{\max}^0(f^1_-) \leq b_{\max}^1(f^1_-)$, the excess-demand in {\sc
    Low} is less than that of {\sc High}. Since the excess demand in {\sc
    High} becomes non-positive at $t = f_1$ (stopping condition), we conclude that excess-demand
  in {\sc Low} will become non-positive at time $t = f^1$ so that {\sc Low}
  will stop at that time, again a contradiction.

  We thus have $q^1 \leq q^0 < f^0 \leq f^1$ and $b_*^0(t) = b_{\max}^0(t)
  \leq b_*^1(t) = b_{\max}^1(t)$ in the interval $q^0 \leq t < f^0$.
  Applying Claim~\ref{cl_rate_bmax_3} and Equation~\ref{eq_utility_db},
\begin{eqnarray*}
  u^0(q^0,f^0_-) =  \int_{q^0}^{f^0_-}
  -\frac{\left(v_*-p(t)\right)}{p(t)}\frac{d}{dt}\left(b_*^0(t)\right)dt \leq
  \int_{q^0}^{f^0_-}
  -\frac{\left(v_*-p(t)\right)}{p(t)}\frac{d}{dt}\left(b^1_*(t)\right) dt =
  u^1(q^0,f^0_-)  
\end{eqnarray*}  

Since auction {\sc Low} stops at time $f^0$ and Alice $\in C^0(f^0_-)$, we
can bound the utility of Alice from the final one shot allocation in {\sc
  Low} as $$u^0(f^0) \leq \frac{(v_* - p(f^0))}{p(f^0)} b_*^0(f^0_-)$$

Assume $v_* > p(f^0)$, else we are already done. In this case, since Alice
$\in C^0(f^0_-) \cap C^1(f^0_-)$, we have $$S^0(f^0_-) - \sum_{i \neq
  \mbox{Alice}} D_i^0(f^0_-) = S^1(f^0_-) - \sum_{i \neq \mbox{Alice}}
D_i^1(f^0_-) = 0$$ Following the proof of Claim~\ref{cl_flip}, we
have $b_i^0(f^0_-) \leq b_i^1(f^0_-)$ for all bidders $i$ with $v_i =
p(f^0)$. Therefore, 
$$ S^0(f^0_-) - \sum_{i \neq \mbox{Alice}, v_i > p(f^0) } D_i^0(f^0_-) \leq
S^1(f^0_-) - \sum_{i \neq \mbox{Alice}, v_i > p(f^0) } D_i^1(f^0_-) $$ 
Since auction {\sc Low} stops at time $f^0$, we have:
$$ \frac{b_*^0(f^0_-))}{p(f^0_-)} \leq S^0(f^0_-) - \sum_{i \neq
  \mbox{Alice}, v_i > p(f^0) } D_i^0(f^0_-) \leq S^1(f^0_-) - \sum_{i \neq
  \mbox{Alice}, v_i > p(f^0) } D_i^1(f^0_-)  $$ 
 Thus, by Lemma~\ref{lem:adjust}, in auction {\sc High}, Alice gets at least
$\frac{b_*^0(f^0_-)}{p(f^0)}$ fraction of the item at price $p(f^0)$, and hence:
$$ u^1(f^0,f^1) \geq \frac{(v_* - p(f^0))}{p(f^0)} b_*^0(f^0_-) \geq u^0(f^0)$$

Therefore, $u^0 = u^0(q^0,f^0_-) + u(f^0) \leq u^1(q^0,f^0_-) +
u^1(f^0,f^1) = u^1$. This completes the proof. 
\end{proof}

\subsubsection{Case 2: $b^1_{\max}(q^0) < b^0_{\max}(q^0)$}
We will now prove Theorem~\ref{th_MAIN} for the case when $b^1_{\max}(q^0) <
b^0_{\max}(q^0)$; this will complete its proof assuming $q^0 < f^0$.  

We now show a sequence of claims bounding the utility obtained in various
phases of the auction.

\begin{claim}
\label{cl_lower_budget}
If $b^1_{\max}(q^0) < b^0_{\max}(q^0)$, then for all $t \in [q^0,f_{\min})$:
$$u^0(q^0,t) \leq u^1(q^0,t) + \frac{\left(v_* -
  p(q^0)\right)}{p(q^0)}\left\{\left(b^0_{\max}(q^0) -
  b^1_{\max}(q^0)\right)-\left(b^0_{\max}(t) - b^1_{\max}(t)\right)\right\}$$ 
\end{claim}
\begin{proof}
  By Claim~\ref{cl_rate_bmax_2} and Claim~\ref{cl_lower_budget2},
  $\frac{d}{dt}\left(b^1_{\max}(t) - b^0_{\max}(t)\right) \geq 0$.  Now applying
  Equation~\ref{eq_utility_db},

\begin{eqnarray*}
  u^0(q^0,t) - u^1(q^0,t) & = & \int_{q^0}^{t}
  -\frac{\left(v_*-p(t)\right)}{p(t)}\frac{d}{dt}\left(b^0_{\max}(t)\right)dt
  + \int_{q^0}^{t}
  \frac{\left(v_*-p(t)\right)}{p(t)}\frac{d}{dt}\left(b^1_{\max}(t)\right)dt
  \\ 
  & = & \int_{q^0}^{t} \frac{\left(v_*-p(t)\right)}{p(t)}
  \frac{d}{dt}\left(b^1_{\max}(t)-b^0_{\max}(t)\right)dt \\ 
  & \leq & \frac{\left(v_*-p(q^0)\right)}{p(q^0)} \int_{q^0}^{t}
  \frac{d}{dt}\left(b^1_{\max}(t)-b^0_{\max}(t)\right)dt \\ 
  & = & \frac{\left(v_*-p(q^0)\right)}{p(q^0)} \left[b^1_{\max}(t) -
  b^0_{\max}(t)\right]_{q^0}^{t}  
\end{eqnarray*}
The claim follows.
\end{proof}

\begin{claim}
\label{cl_q1q0}
If $b^1_{\max}(q^0) < b^0_{\max}(q^0)$, then:
$$ u^1(q^1,q^0) \geq \frac{\left(v_* - p(q^0)\right)}{p(q^0)}
\left(b^0_{\max}(q^0) - b^1_{\max}(q^0)\right) $$ 
\end{claim}
\begin{proof}
  Consider auction {\sc High}. Alice starts to clinch at time $q^1$. As the
  price increased from $p(q^1)$ to $p(q^0)$, her budget decreased by an
  amount $ b^1_{\max}(q^1) - b^1_{\max}(q^0) $.  The price was always less
  than $p(q^0)$ during this interval; thus she gets at least
  $(1/p(q^0))\left(b^1_{\max}(q^1) - b^1_{\max}(q^0)\right)$ fraction of the
  item at an average unit price that is at most $p(q^0)$. We get $$
  u^1(q^1,q^0) \geq \frac{\left(v_* - p(q^0)\right)}{p(q^0)}
  \left(b^1_{\max}(q^1) - b^1_{\max}(q^0)\right) $$ By definition,
  $b^1_{\max}(q^1) = B^1_* > B^0_* = b^0_{\max}(q^0)$, and the claim is
  proved.
\end{proof}

\begin{claim}
\label{cl_LOWER_BUDGET}
If $b^1_{\max}(q^0) < b^0_{\max}(q^0)$, then for all $t \in [q^0,f_{\min})$: 
$$u^0(q^0,t) \leq u^1(q^1,t) -
\frac{\left(v_* - p(t)\right)}{p(t)}\left(b^0_{\max}(t) - b^1_{\max}(t)\right)$$
\end{claim}
\begin{proof}
Applying Claim~\ref{cl_lower_budget},~\ref{cl_q1q0}, we  get
\begin{eqnarray*}
u^0(q^0,t) & \leq & u^1(q^0,t) + u^1(q^1,q^0) - \frac{\left(v_* -
  p(q^0)\right)}{p(q^0)}\left(b^0_{\max}(t) - b^1_{\max}(t)\right) \\ 
& = & u^1(q^1,t) - \frac{\left(v_* - p(q^0)\right)}{p(q^0)}\left(b^0_{\max}(t)
- b^1_{\max}(t)\right) \\ 
& \leq & u^1(q^1,t) - \frac{\left(v_* - p(t)\right)}{p(t)}\left(b^0_{\max}(t) -
b^1_{\max}(t)\right) 
\end{eqnarray*}
\end{proof}

\begin{lemma}
\label{lem:more}
If $b^1_{\max}(q^0) < b^0_{\max}(q^0)$, then $u^0 \le u^1$. 
\end{lemma}
\begin{proof}
  Similar to the proof of Lemma~\ref{lem:less}, it can be shown that $f^1
  \leq f^0$. Putting $t = f^1_-$ in Claim~\ref{cl_LOWER_BUDGET}, 
\begin{equation}
\label{eq_lm_lower_budget_1}
u^0(q^0,f^1_-) + \frac{\left(v_* - p(f^1)\right)}{p(f^1)} b^0_{\max}(f^1_-)
\leq u^1(q^1,f^1_-) + \frac{\left(v_* - p(f^1)\right)}{p(f^1)}
b^1_{\max}(f^1_-) 
\end{equation}

If $p(f^1) = v_*$, then $f^0 = f^1$ and Alice receives zero utility from
the final one-shot allocations in both the auctions. Note that by
Claim~\ref{cl_lower_budget2}, we have $b^1_{\max}(f^1_-) \leq
b^0_{\max}(f^1_-)$. Thus, $u^0 = u^0(q^0,f^1_-) \leq u^1(q^1,f^1_-) = u^1$
and the lemma is true.  

\medskip Now suppose $p(f^1) < v_*$. Alice's utility from the final
one-shot allocation in {\sc High} is given by: 
$$ u^1(f^1) = \frac{\left(v_* - p(f^1)\right)}{p(f^1)} b^1_{\max}(f^1_-) $$
On the other hand, in {\sc Low}, during the time interval $f^1 \leq t \leq
f^0$, Alice can get at most $\frac{b^0_{\max}(f^1_-)}{p(f^1)}$ fraction of
the item at an average unit price that is at least $p(f^1)$. Thus,
$$u^0(f^1,f^0) \leq \frac{\left(v_* -
    p(f^1)\right)}{p(f^1)}b^0_{\max}(f^1_-)$$ 
Adding this to Equation~\ref{eq_lm_lower_budget_1}, we get
$$ u^0 = u^0(q^0,f^1_-) + u^0(f^1,f^0) \leq u^1(q^1,f^1_-) + u^1(f^1) = u^1 $$
This completes the proof.
\end{proof}

The proof of Theorem~\ref{th_MAIN} for the case when $q^0 < f^0$ now follows from Lemmas~\ref{lm_main},~\ref{lem:less}
and~\ref{lem:more}.

\subsection{The Special Case: Alice Never Enters $C(t)$  in Auction {\sc Low}, that is $q^0 = f^0$}
\label{sec:corner}
In this section, we prove Theorem~\ref{th_MAIN} when $q^0 = f^0$, that is, in auction {\sc Low}, Alice receives all her utility from the final one shot allocation in Step (I) of Figure~\ref{fig:clinch}. We will consider three mutually exclusive and exhaustive cases corresponding respectively to {\sc Low} stopping: (i) before any bidder starts clinching; (ii) after some bidder starts clinching, but before any bidder starts clinching in {\sc High}; and (iii) after some bidder starts clinching in {\sc High}. We first show the following claim which gives a closed form expression for the utility gained by Alice in {\sc Low}.

\begin{claim}
\label{cl:ulow}
In auction {\sc Low}, Alice only receives a one shot allocation of $\frac{B_*^0}{p(q^0)}$ at price $p(q^0)$, and her utility is given by  $u^0 = \frac{\left(v_*-p(q^0)\right)}{p(q^0)}B_*^0$. Furthermore, in this case,  
$$\sum_{i \ : \ v_i = p(q^0)} b^0_i(q^0_-) \geq b^0_{\max}(q^0_-)$$. 
\end{claim}
\begin{proof}
  Consider auction {\sc Low}. Since $q^0 = f^0$, Assumption~\ref{as_p(q0)<v*} implies $p(f^0) < v_*$, so
  that by Lemma~\ref{cor_Bi}, Alice's budget is extracted completely at
  price $p(q^0)$. The first part of the claim follows. 

To see the second part, first note that Supply Invariant holds just before the auction stops. At time $q^0_-$, from the perspective of the active bidder with  highest remaining budget ($b^0_{\max}(q^0_-)$), total demand of other active bidders is no less than the available supply ($S^0(q^0_-)$). In other words:
$$ \sum_{i \in A(q^0_-)} \frac{b^0_i(q^0_-)}{p(q^0)} \geq S^0(q^0_-) + \frac{b^0_{\max}(q^0_-)}{p(q^0)} $$
At time $q^0$, the auction stops because total demand of all the active bidders is no more than available supply, so that $ \sum_{i \in A(q^0)} \frac{b^0_i(q^0_-)}{p(q^0)} \leq S^0(q^0_-) $. Thus, total demand of active bidders drop by at least $\frac{b^0_{\max}(q^0_-)}{p(q^0)}$ as time changes from $q^0_-$ to $q^0$. This abrupt decrease in total demand is caused by the set of exiting bidders (that is, bidders with $v_i = p(q^0)$). Therefore, we get: $\sum_{i \ : \ v_i = p(q^0)} b^0_i(q^0_-) \geq b^0_{\max}(q^0_-)$, completing the proof.
\end{proof}

\subsubsection{Case 1:  $y^0 = q^0$}
We first consider the case where {\sc Low} stops before any bidder starts clinching. We have $y^0 = q^0 = f^0$. Using an argument similar to the proof of Claim~\ref{cl_very_begin}, it can be shown that {\sc Low} starts clinching no later than {\sc High}, that is, $y^0 \leq y^1$. Furthermore, just before {\sc Low} starts clinching (at time $y^0_-$), the remaining budget of every active bidder equals her reported budget. In particular, every bidder $i$ other than Alice has the same remaining budget across the two auctions, that is, $b_i^0(y^0_-) = b_i^1(y^0_-)$. For Alice, $b^1_*(y^0_-) = b^0_*(y^0_-) + \Delta$. Also note that $S^0(y^0_-) = S^1(y^0_-) = 1$. Since auction {\sc Low} stops at time $y^0$, we must have:
$$ \sum_{i \ : \ v_i > p(y^0)} \frac{b^1_i(y^0_-)}{p(y^0)} = \frac{\Delta}{p(y^0)} + \sum_{i \ : \ v_i > p(y^0)} \frac{b^0_i(y^0_-)}{p(y^0)} \leq \frac{\Delta}{p(y^0)} + S^0(y^0_-) = \frac{\Delta}{p(y^0)} + S^1(y^0_-) $$ 
Comparing the LHS and the RHS, we see that in auction {\sc High}, 
$$ \frac{B_*^0}{p(y^0)} + \sum_{i \ : \ v_i > p(y^0), i \neq \mbox{Alice}} \frac{b^1_i(y^0_-)}{p(y^0)} \leq S^1(y^0_-) $$ 
Thus, Alice receives at least $\frac{B_*^0}{p(y^0)}$ fraction of the item at unit price $p(y^0)$ in auction {\sc High}. Since $y^0 = q^0$, Claim~\ref{cl:ulow} implies her utility from {\sc High} is no less than her utility from {\sc Low}.

\subsubsection{Case 2:  $y^0 < q^0 \leq y^1$}
We next consider the case where {\sc Low} stops after some bidder starts clinching, but before any bidder starts clinching in {\sc High}. Let $A$ stands for ``Alice". Suppose $y^0 < q^0 \leq y^1$. Since {\sc Low} stops at $q^0$,
\begin{eqnarray*} 
\sum_{i \ : \ v_i > p(q^0), i \neq A} \frac{b^0_i(p(q^0_-))}{p(q^0)} + \frac{B_*^0}{p(q^0)} \leq S^0(q^0_-) \\
\end{eqnarray*}
For all bidders $i$ with $v_i > p(q^0), i \neq A$, we have $b^1_i(q^0_-) =
B_i$, otherwise $y^1 < q^0$. Since in {\sc Low}, clinching bidders
clinched at price at most $p(q^0)$, we have
\begin{eqnarray*}
  \sum_{i \ : \ i \in C^0(q^0_-), v_i > p(q^0)} \frac{B_i - b_i^0(q^0_-)}{p(q^0)} & \leq & 1 - S^0(q^0_-) \\
\Rightarrow  \sum_{i \ : \ i \in C^0(q^0_-), v_i > p(q^0)} \frac{b^1_i(q^0_-)}{p(q^0)} = \sum_{i \ :
    \ i \in C^0(q^0_-), v_i > p(q^0)} \frac{B_i}{p(q^0)} &\leq & 1 + \sum_{i \ : \ i \in
    C^0(q^0_-), v_i > p(q^0)} \frac{b_i^0(q^0_-)}{p(q^0)} - S^0(q^0_-) \\ 
\end{eqnarray*}
For all bidders $i$ with $v_i > p(q^0), i \neq A, i \notin C^0(q^0_-)$, we
have $b^1_i(q^0_-) = b^0_i(q^0_-) = B_i$. It follows that
\begin{eqnarray*}
  \sum_{i \ : \ v_i > p(q^0), i \neq A} \frac{b^1_i(q^0_-)}{p(q^0)} +
  \frac{B_*^0}{p(q^0)} \leq 1 + \sum_{i \ : \ v_i > p(q^0), i \neq A}
  \frac{b^0_i(q^0_-)}{p(q^0)} + \frac{B_*^0}{p(q^0)} - S^0(q^0_-) \leq 1 =
  S^1(q^0_-) 
\end{eqnarray*}
Therefore, by Lemma~\ref{lem:adjust}, Alice clinches at least $B_*^0/p(q^0)$ quantity in {\sc High} at
price $p(q^0)$, so that Theorem~\ref{th_MAIN} holds.

\subsubsection{Case 3: $y^0 \leq y^1 < q^0$}
We finally consider the case where {\sc Low} stops after some bidder starts clinching in both auctions. We have $y^0 \leq y^1 < q^0 = f^0 < f^1$ (see
Assumption~\ref{as_stop_time}). Since the second
inequality is strict, we get $C^0(q^0_-), C^1(q^0_-) \neq \emptyset$.  

Following an argument exactly similar to the one outlined in Section~\ref{sec:struct}, we have $b_{\max}^1(y^1) \le
b^0_{\max}(y^1) + \Delta$, and whenever
$b^1_{\max}(t) \ge b^0_{\max}(t)$, the former reduces at a larger rate. 
  Therefore:
  \begin{equation}
  \label{eq:bmq0}
  b_{\max}^1(q^0) \le b^0_{\max}(q^0_-) + \Delta
  \end{equation}

  \medskip We will first show that $q^1 < q^0$, else Theorem~\ref{th_MAIN}
  is true. Suppose $q^1 \ge q^0$. By Claim~\ref{cl:ulow}, Alice gets a one
  shot allocation of $B_{*}^0/p(q^0)$ at stopping price $p(q^0)$. We will
  show that Alice will also get at least $B_*^0/p(q^0)$ at the same price
  in {\sc High}. Now, since $q^0 \le q^1$ and $q^0 < f^1$, we must have:
\begin{eqnarray*}
  D^1_{-\mbox{Alice}}(q^0) + \frac{B_*^1}{p(q^0)}  = \sum_k D^1_k(q^0) & =
  & S^1(q^0) + \frac{b^1_{\max}(q^0)}{p(q^0)}  \\ 
  \Rightarrow \ \ S^1(q^0) - D_{-\mbox{Alice}}^1(q^0) & = &  -
  \frac{b^1_{\max}(q^0)}{p(q^0)} + \frac{B_*^1}{p(q^0)} 
\end{eqnarray*}
Since auction {\sc Low} stops at time $q^0$, by Claim~\ref{cl:ulow}, we
must have $$\sum_{i \ : \ v_i = p(q^0)} b^0_i(q^0_-) \geq
b^0_{\max}(q^0_-)$$ Note that a bidder with valuation equal to $p(q^0)$ can
never be in $C^1(q^0)$, otherwise we will have $q^0 = f^1$, a
contradiction. Also note by Claim~\ref{cl:ulow} that Alice does not have
valuation equal to $p(q^0)$. Thus, for all bidders $i$, if $v_i = p(q^0)$,
then $b_i^1(q^0) = B_i^1 = B_i^0 \geq b_i^0(q^0_-)$. That is,
$$ \sum_{i \ : \ v_i = p(q^0)} b^1_i(q^0) \geq \sum_{i \ : \ v_i = p(q^0)}
b^0_i(q^0_-) \geq b^0_{\max}(q^0_-) $$ 
It follows that
$$ S^1(q^0) - \sum_{j \in A(q^0), j \neq \mbox{Alice}} D^1_j(q^0)\geq -
\frac{b^1_{\max}(q^0_-)}{p(q^0)} + \frac{B_*^1}{p(q^0)} +
\frac{b_{\max}^0(q^0_-)}{p(q^0)} \geq \frac{B_*^1 - \Delta}{p(q^0)} =
\frac{B_*^0}{p(q^0)}$$ 
The final inequality follows from Equation (\ref{eq:bmq0}). Therefore, by Lemma~\ref{lem:adjust},
Alice clinches at least $B_*^0/p(q^0)$ at price $p(q^0)$ in {\sc High} to
maintain the Supply Invariant, and Theorem~\ref{th_MAIN} is true.

\medskip
Therefore, if Theorem~\ref{th_MAIN} is not  already true, we must have:
$y^0 \leq y^1 \leq q^1 < q^0 = f^0 < f^1$. In particular,  Alice is
clinching in {\sc High} during the time interval $q^1 \leq t <
f^1$. Furthermore, we have $C^0(q^0_-), C^1(q^0) \neq \emptyset$ and Alice $\in
C^1(q^0)$. Similar to the argument above, we must have: 
$$ \sum_{i \ : \ v_i = p(q^0)} b_i^1(q^0) \geq \sum_{i \ : \ v_i = p(q^0)}
b_i^0(q^0_-) \geq b_{\max}^0(q^0_-) $$ 
Since Alice $\in C^1(q^0)$, we have:
\begin{eqnarray*}
 S^1(q^0) - D^1_{-\mbox{Alice}}(q^0)  & = & 0 \\
\Rightarrow \ \ S^1(q^0) - \sum_{j \in A(q^0), j \neq \mbox{Alice}}D_j^1(q^0) & \geq & \frac{b_{\max}^0(q^0_-)}{p(q^0)}
\end{eqnarray*}
Therefore, in {\sc High}, by Lemma~\ref{lem:adjust}, Alice gets at least
$\frac{b_{\max}^0(q^0_-)}{p(q^0)}$ fraction at unit price $p(q^0)$. Also
note that Alice reduced her budget from $B_*^1$ to $b^1_{\max}(q^0)$ during
the time interval $q^1 \leq t < q^0$. Thus, in this time interval, she
clinched at least $\frac{\left(B_*^1 - b^1_{\max}(q^0)\right)}{p(q^0)}$ at
an average unit price that is at most $p(q^0)$. Thus, we conclude: 
\begin{eqnarray*}
u^1 & \geq & \frac{v_* - p(q^0)}{p(q^0)} \left(B_*^1 - b_{\max}^1(q^0) +
  b_{\max}^0(q^0_-)\right) \\ 
& \geq & \frac{v_* - p(q^0)}{p(q^0)} \left(B_*^1 - \Delta\right) = u^0
\end{eqnarray*}
The final inequality follows from Equation (\ref{eq:bmq0}). This implies
Theorem~\ref{th_MAIN}.


\begin{thebibliography}{10}

\bibitem{abrams}
Z.~Abrams.
\newblock Revenue maximization when bidders have budgets.
\newblock In {\em SODA '06: Proceedings of the seventeenth annual ACM-SIAM
  symposium on Discrete algorithm}, pages 1074--1082, 2006.

\bibitem{ausubel}
L.~Ausubel.
\newblock An efficient ascending-bid auction for multiple objects.
\newblock {\em American Economic Review}, 94(5):1452--75, December 2004.

\bibitem{borgs}
C.~Borgs, J.~Chayes, N.~Immorlica, M.~Mahdian, and A.~Saberi.
\newblock Multi-unit auctions with budget-constrained bidders.
\newblock In {\em EC '05: Proceedings of the 6th ACM conference on Electronic
  commerce}, pages 44--51, 2005.

\bibitem{pino}
S.~Brusco and G.~Lopomo.
\newblock Simultaneous ascending bid auctions with privately known budget
  constraints.
\newblock {\em Journal of Industrial Economics}, 56(1):113--142, 2008.

\bibitem{shuchi}
S.~Chawla, J.~D. Hartline, and R.~D. Kleinberg.
\newblock Algorithmic pricing via virtual valuations.
\newblock In {\em ACM Conference on Electronic Commerce}, pages 243--251, 2007.

\bibitem{nisan}
S.~Dobzinski, R.~Lavi, and N.~Nisan.
\newblock Multi-unit auctions with budget limits.
\newblock In {\em FOCS}, pages 260--269, 2008.

\bibitem{Goldberg04}
A.~V. Goldberg, J.~D. Hartline, and A.~Wright.
\newblock Competitive auctions and digital goods.
\newblock In {\em SODA}, pages 735--744, 2001.

\bibitem{timr}
J.~D. Hartline and T.~Roughgarden.
\newblock Optimal mechanism design and money burning.
\newblock In {\em STOC}, pages 75--84, 2008.

\bibitem{laffont}
J.-J. Laffont and J.~Robert.
\newblock Optimal auctions with financially constrained buyers.
\newblock {\em Economics Letters}, 52:181--186, 1996.

\bibitem{maskin}
E.S. Maskin.
\newblock Auctions, development and privatization: {E}fficient auctions with
  liquidity-constrained buyers.
\newblock {\em European Economic Review}, 44:667--681, 2000.

\bibitem{myerson}
R.~B. Myerson.
\newblock Optimal auction design.
\newblock {\em Mathematics of Operations Research}, 6(1):58--73, 1981.

\bibitem{vohra}
M.~Pai and R.~Vohra.
\newblock Optimal auctions with financially constrained bidders.
\newblock {\em Working Paper}, 2008.
\newblock Available at: {\tt
  http://www.kellogg.northwestern.edu/faculty/Vohra/ftp/LR3.pdf}.

\end{thebibliography}
\end{document}